# The filling law: a general framework for leaf shape diversity and its consequences on folded leaves

Etienne Couturier, Sylvain Courrech du Pont, Stéphane Douady

Laboratoire Matières et systèmes complexes (MSC)
UMR 7057 CNRS/Université Paris-Diderot
10 rue Alice Domon & Léonie Duquet
75205 Paris Cedex 13, France

**Abstract**

Leaves are packed in a bud in different ways, being flat, enrolled, or folded, but always filling the whole bud volume. This « filling law » has many consequences, in particular on the shape of growing folded leaves. This is shown here for different types of folding and packing. The folded volume is always part of an ellipsoid, with the veins on the outside rounded face, and the lamina margin on an adaxial plane or axis. The veins on the abaxial side protect the more fragile lamina. The first general consequence of the folds is the presence of symmetries on the leaf shape, but also quantitative relationships between lobes and sinus sizes. For particular geometries, the leaf lamina can be limited by lateral veins, creating spoon-like lobes, or transverse cuts, creating asymmetrical wavy perimeters. A change in the packing between cultivars induces the corresponding change in the leaf shape. Each particular case shows how pervasive are the geometrical consequences of the filling law.

## 1 - Introduction

Leaves present many different shapes. They can be grouped in various categories, for instance simple, with lobes (palmate), with leaflets (compound), or dissected with holes. But even on one single plant, the leaf shape can vary, sometimes strongly, leading to heterophilly. Despite this diversity, some common features are intuitively guessed between leaves even from very different phylogenetic origins. For instance, folded palmated leaves seem to share a common regulation of the respective lobes dimensions.

Leaf first appears as primordia, little bumps around the centre of the apical meristem,. Very quickly a primordium expands ortho-radially, as a surface wrapping around the stem axis. From the beginning they present a fundamental asymmetry: the side turned toward the stem axis (adaxial) will become the smooth and shiny upper side of the leaf turned toward the light; the other side, turned toward outside (abaxial), present hairs and veins protruding and will become the lower side of the leaf.

The lobes appearance and development can first be considered as a reiteration. Lobes are secondary primordia protruding from the margin of his surface, as a reiteration of the leaf primordia itself. At gene's level, it has been shown for coumpound leaves that the same genes are expressed in both case: the gene CUC is expressed at the separation between the primordia and the meristem, and at the separation between the leaflets [Blein et al (2009)].

As a development of a lobe is necessary linked with the development of a vein, lobes and veins have the same hierarchy. Each main lobe corresponds to a major vein ending at its tip. Similarly, secondary veins branch out from them, possibly developing secondary lobes, and so on and so forth.

If lobes and veins initiation begins to be understood by developmental biologists, what regulates their final size remains unclear. We have shown in a previous article that for palmated leaves, a key feature of this regulation, is that these leaves develop folded inside the bud [Couturier et al. (2009)]. The fact that it is common for leaves to grow folded has been noticed early, but overlooked since the XIXth century [Clos (1870)]. The only case, which has been studied in detail is the palm leaves, a monocotyledon [Dengler et al. (1982)]. For dicotyledons, there are very few studies on leaf folds [Kobaiashi et al. (1998)].

All the buds from different plants, taken in the general meaning of the compact organisation of leaves around the meristem apex (and not only immature leaves protected by scales [Bell (1991)]), are very well organized. The bud internal space is perfectly filled by the successive immature leaves, which all occupy a volume with the same shape, but of decreasing size. The shape, which is occupied by a leaf, does not depend much on the species: it is roughly a quarter of an ellipsoid: the curved face is the abaxial leaf side and the large plane face is the adaxial one (the small one being essentially the base of the leaf).

If all the leaves occupy the same volume in the bud, they have different ways of filling it. What is surprising, and never noticed before, is that the corresponding leaf shape is exactly the one filling the bud volume once folded. We call this the "filling law". We will show that this law is very strong and general among the species and present its consequences on leaf shape, in particular for folded leaves.

## 2 - Results

### *2.1 - Compactness of the bud*

Many dicotyledon leaves of various phylogenetic origins are folded during their development. Even if these plants follow different phyllotactic patterns, their buds have a similar organisation. They fill a the part of ellipsoid volume, with the veins on the outside rounded part of the ellipsoid, and the lamina, always as locally flat as possible, folded inside. The leaf margin lays on the flat border. The difference in organisation just comes from the different objects in the bud delimiting the internal border. In the *Acer platanus* type, the object is the margin of the opposite folded leaf (Figure 1a). In the *Morus platanifolium* type, the objects are the stipula of the next leaf that are joined in a closed envelope that protects the next younger leaf (Figure 1b). In the *Pelargonium cuculatum* type, it corresponds to the bottom part of an older leaf (Figure 1c).

In a last fourth bud organisation type, typical of tropical climate, the leaf is globally wrapped around itself, so that the inside delimitation is not another leaf but the leaf itself. In this way each leaf behaves as an autonomous bud (Figure 1d-d'), where the leaf margin is limited on a central axis. These four organisations enable a perfect filling and tilling of the bud internal space (Figure 2a-b-c).

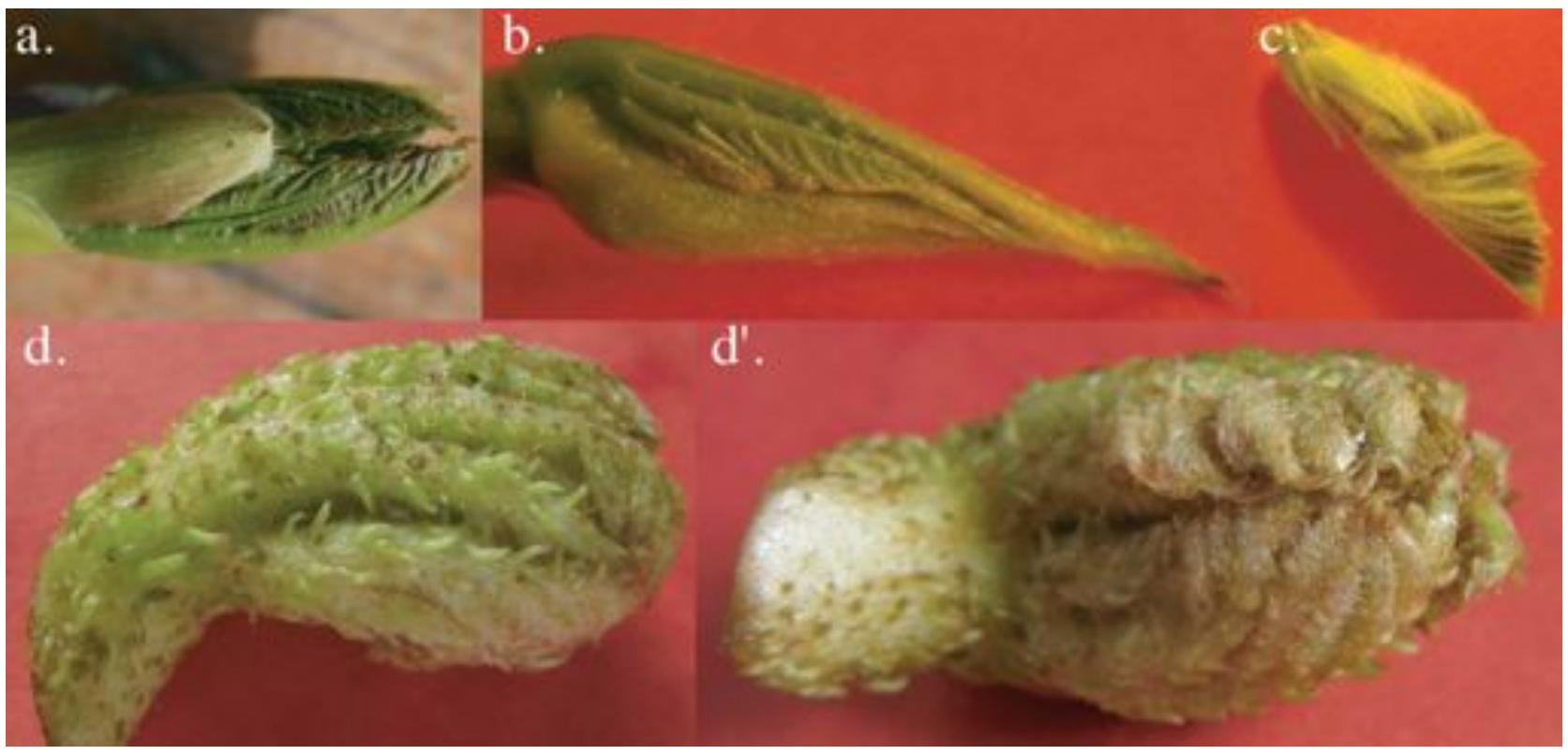

Figure 1 : A similar structure in different kind of buds. The external part is always constituted by the veins, and the lamina is folded inside. a. *Acer pseudoplatanus* bud. The folded leaf margin lays on a plane which corresponds to the opposite leaf margin. b. *Morus platanifolium* leaf. The folded leaf margin lays on a plane which corresponds to the stipule which envelopes a smaller bud contained in the bud itself. c. *Pelargonium cuculatum*. The folded leaf margin lays on a plane which corresponds to the older leaf surface. d. *Gunera manicata*. View of one side of the leaf. d'. View of the face of the same leaf. The leaf constitutes its own bud, and the lamina is delimited by the central axis (as seen in Fig. 2 c).

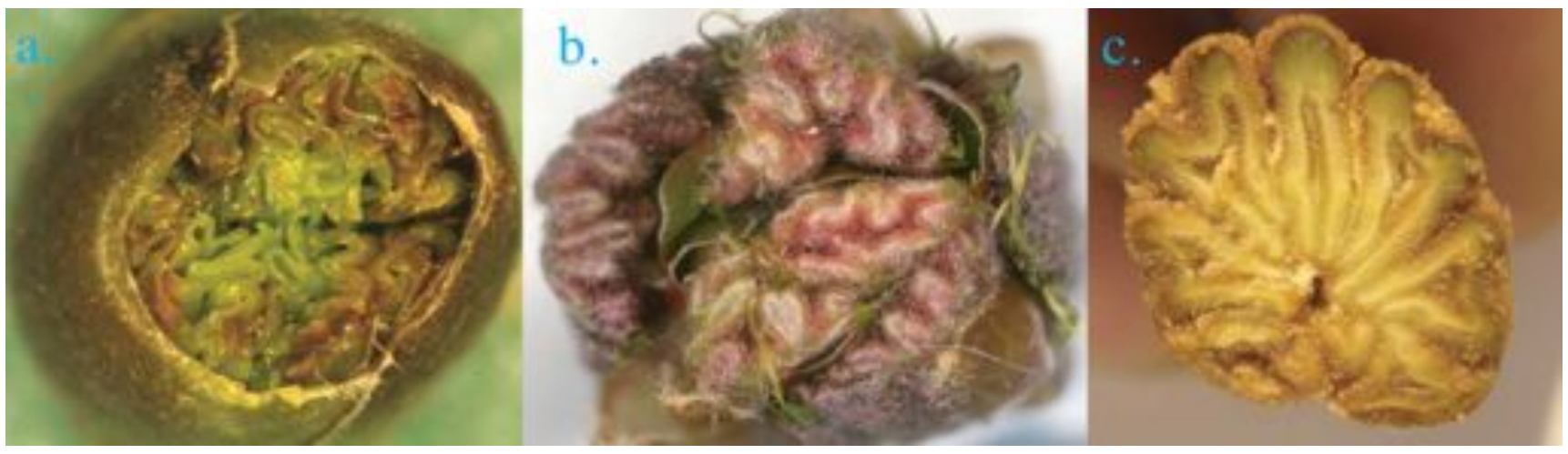

Figure 2 : Three cuts of the three different bud kinds : a perfect and rational fulling of the bud. . a. Opposite phyllotaxy, *Acer pseudoplatanus*. b. Spiral phyllotaxy, *Quercus Rubra*. c. Autoclosing leaf, *Tetrapanax papyriferum*.

*2.2 - Basic perimeter symmetry*

In all the types, the minimum consequence of this organisation is that, even if the actual folding occurs in 3D with thick veins, two consecutive main lobes margin are folded on the same line. As on both side of a fold the leaf margin lays on the same line, each fold corresponds to the symmetry axis of the perimeter of the unfolded leaf, either for a sinus or for a lobe. The abaxial folds, which correspond to a lobe, follow a main vein (Figure 3a). The adaxial folds, which correspond to a sinus, are in the lamina (Figure 3b). They correspond to the zone between two main veins, where the smaller secondary veins join (Figure 3c). We call them antifolds, or “anti-veins”.

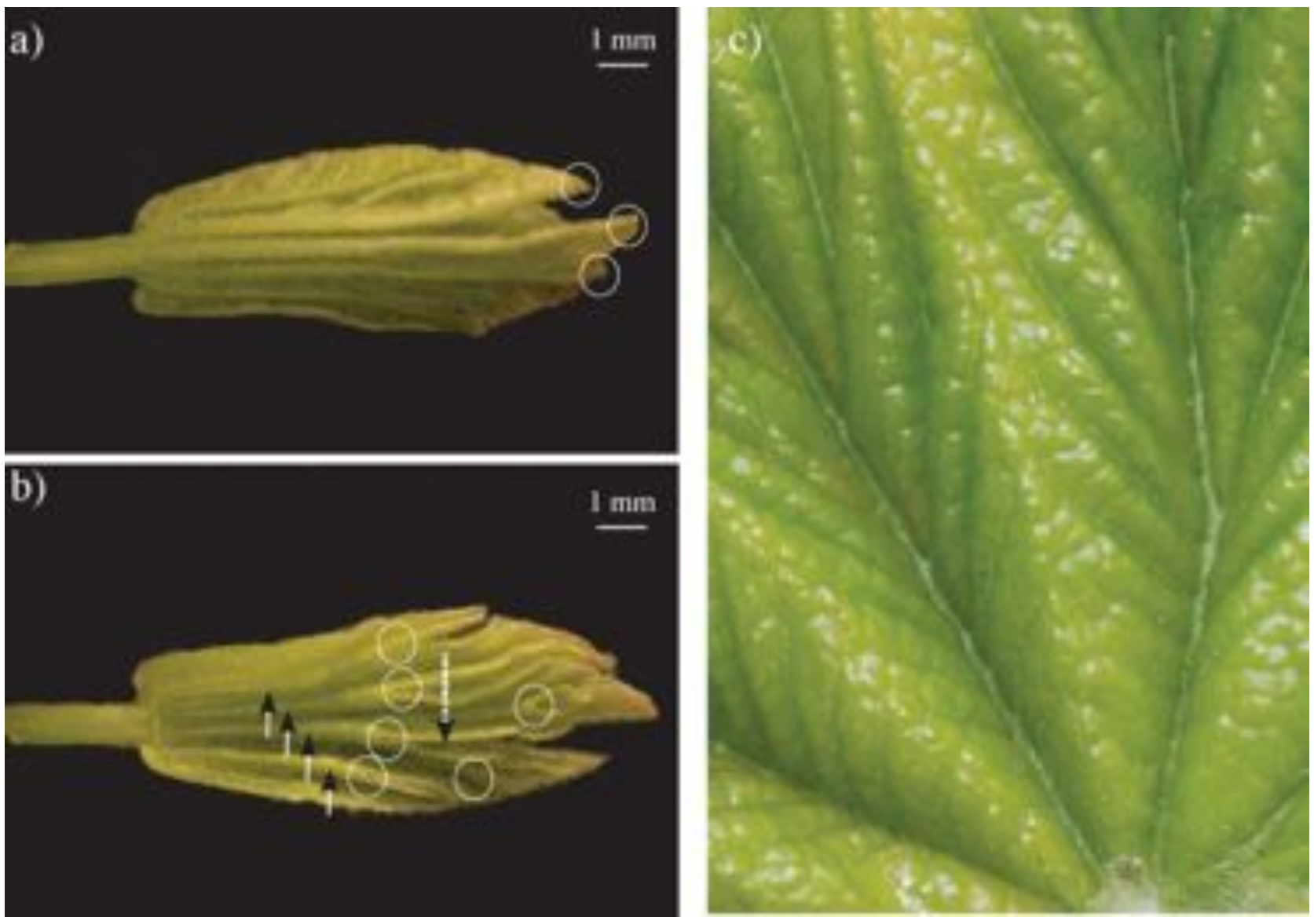

Figure 3: Folded immature leaf extracted from a bud of *Acer campestre*. a The abaxial side shows the anticlinal folds running along veins and ending at peaks (circles). b The adaxial side shows the synclinal folds (arrows) running along (immaterial) anti-veins and ending at sinuses (circles). Peaks and sinuses stand in the contact plane of the pair of leaves (figure 1), but while peaks are at the extreme of this contact surface, sinuses are inside. C. An antifold of *Acer pseudoplatanus*. Only the last order of veins, emanating from surrounding main veins, join along the antifold.

As the symmetry of the leaf margin around a fold or antifold is a packing geometrical constrain, it works for all folded species independently of the phylogeny. (Figure 4).

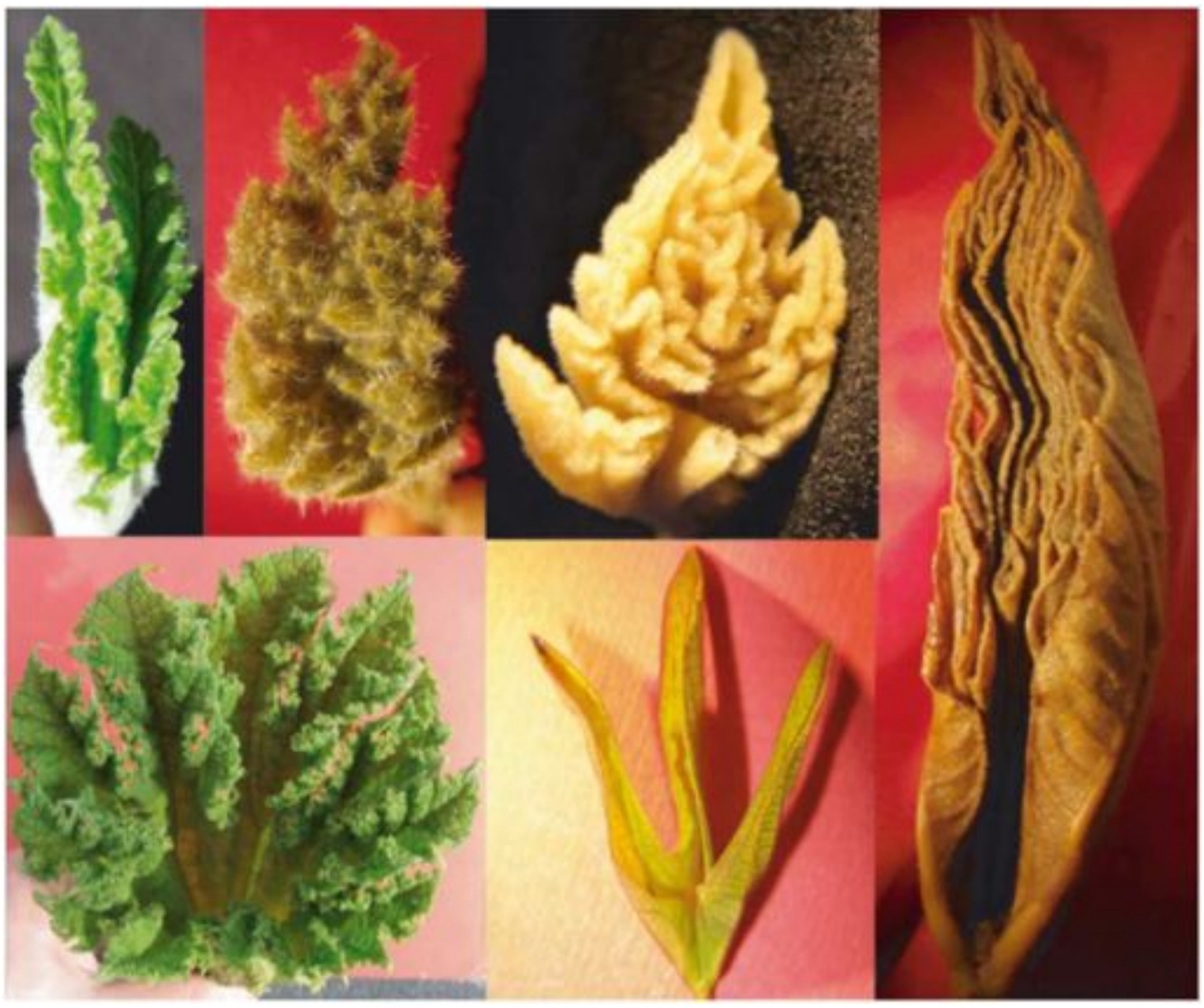

Figure 4: Front view of different fold and cut leaves . *a. Ribes Nigrum. b. Pelargonium cuculatum. c. Malva Sylvestris. d. Tetrapanax papyrifer. e. Gunnera manicata. f. Passiflora.* They are of different phyllogeny : a is a *Saxifragales*, b a *Geraniales*, c a *Sapindales*, d. an *Apiales*, e. a *Gunnerales* and f. a *Malpighiales*.

This symmetry property is preserved in the mature leaf. The veins, which correspond to the abaxial folds, are the medial axes of the lobe even for mature leaves (Figure 5a,c). The antifolds correspond also to the symmetric of the medial axes of the sinuses (Figure 5b). Because of the preservation of these symmetries during the leaf expansion, the leaf of these species can be refolded even mature (Figure 6). The delimitation of the folded lamina by a external plane is similar to the case of a sheet of paper folded and cut. We thus call this limitation Kirigami (cut-(folded)-paper in Japanese, see Figure 7).

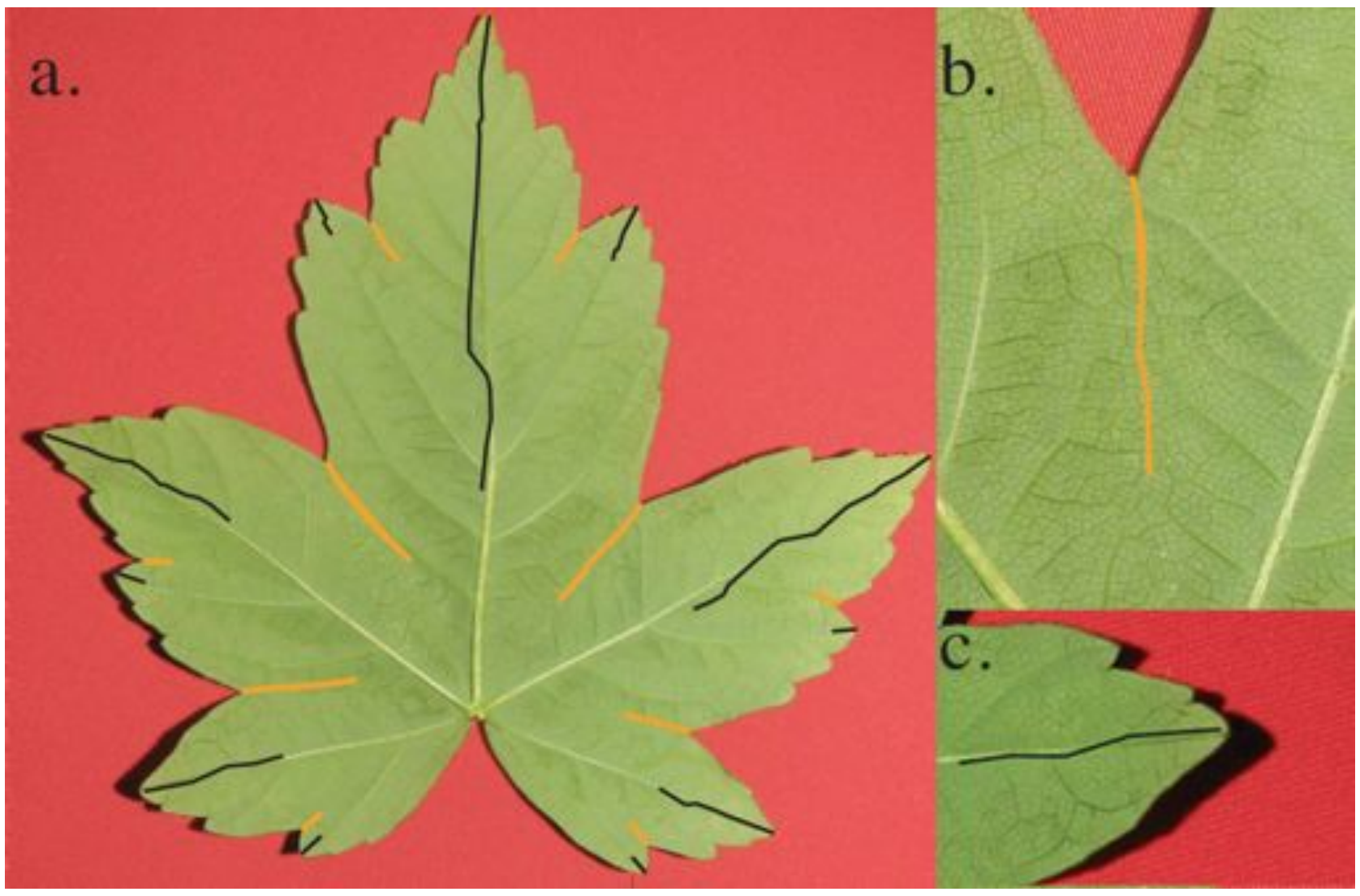

Figure 5 : a. An *Acer pseudoplatanus* leaf. Black lines are medial axis of lobes, derived from the perimeter. Orange lines are the symmetric around the sinus minmum of the medial axis of the sinus b. The orange line corresponds to an antifold. c. The black line corresponds to a fold

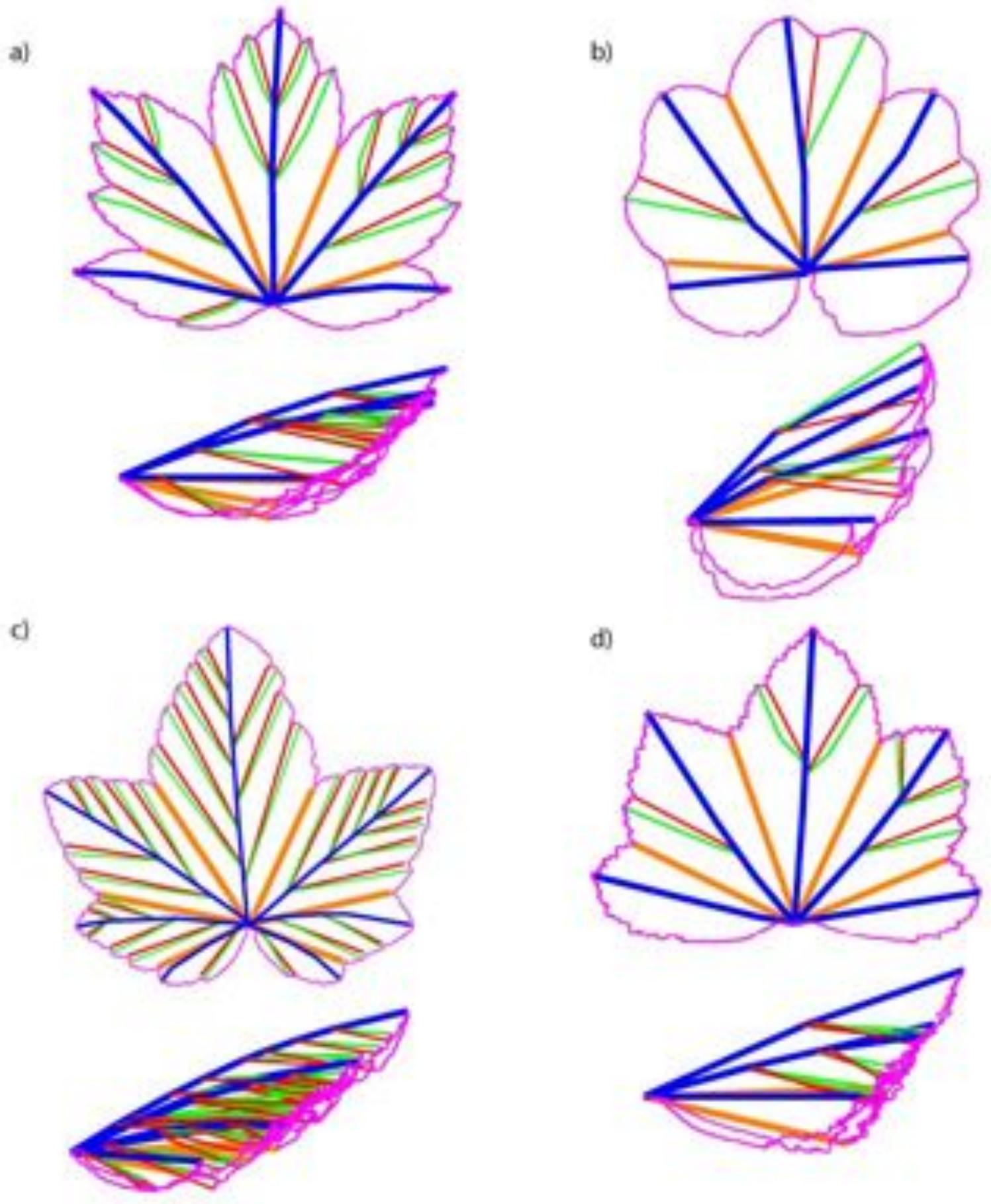

Figure 6 : Numerical folding of some species. a *Acer pseudoplatanus.* b *Gunera manicata.* c *Ribes Nigrum.* d *Malva sylvatica.* Note that the perimeter refolds on a single line even for asymmetric leaves, as in d.

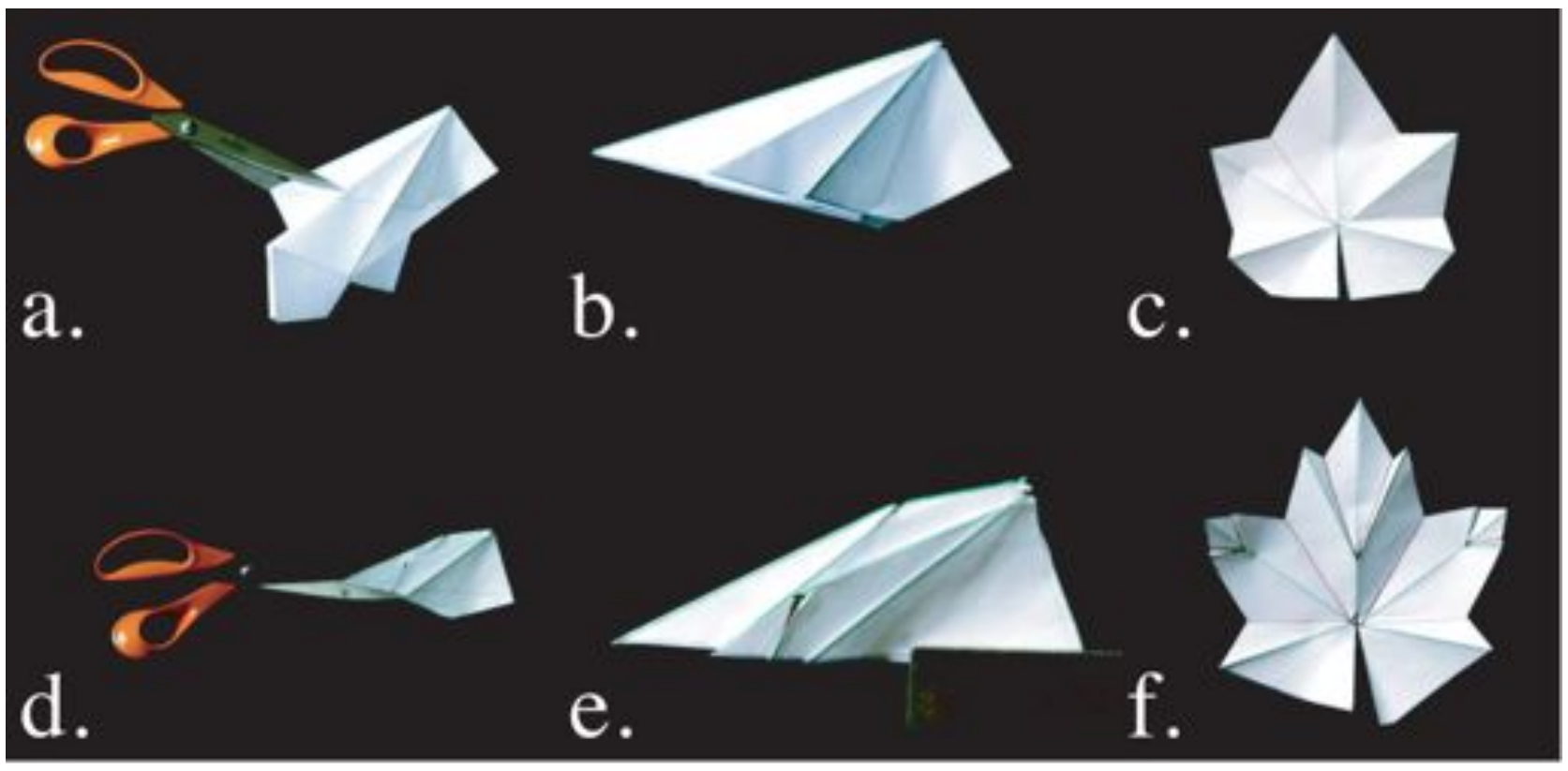

Figure 7: Examples of cut folded paper in the shape of leaves (Kirigami –'cut paper' in japanese) a. Five folds, which originate from the same point (petiole) of a rectangular sheet of paper, to be cut with scissors. The folded margin is complicated. b. Same folded sheet of paper once cut. The folded margin lays on a single line. c. Same sheet once unfolded. Folds correspond to sinuses and lobes. d. Same sheet folded with secondary folds. e. Same sheet once cut f. Same sheet once unfolded. Secondary folds correspond to secondary lobes.

*2.3 - Antifold symmetry*

Leaves have other symmetries originating from them growing folded in a particular way. For all the folded leaves the main veins constitute an outside envelope of contiguous veins around the lamina, protecting it as an armour (Figure 8 a-d). In order for a vein to become contiguous to its neighbouring one, the antifold has to bisect the flat lamina (Figure 8e). For the same reason, in case of more than two veins around an antifold the antifold had to be constituted of pieces of bisectrix of the veins taken two by two (Figure 8f), even if in this case it is no more a globally flat surface. In the common case of an antifold between two secondary veins, it then takes the typical shape of a wedged roof (see fig. 8f'). As the antifolds correspond to the bisectrix of the veins, the sinuses, which are at the end of the antifolds, are on bisectrix of the veins (Figure 8e-f). The final contour of the leaf is strongly constrained by this property.

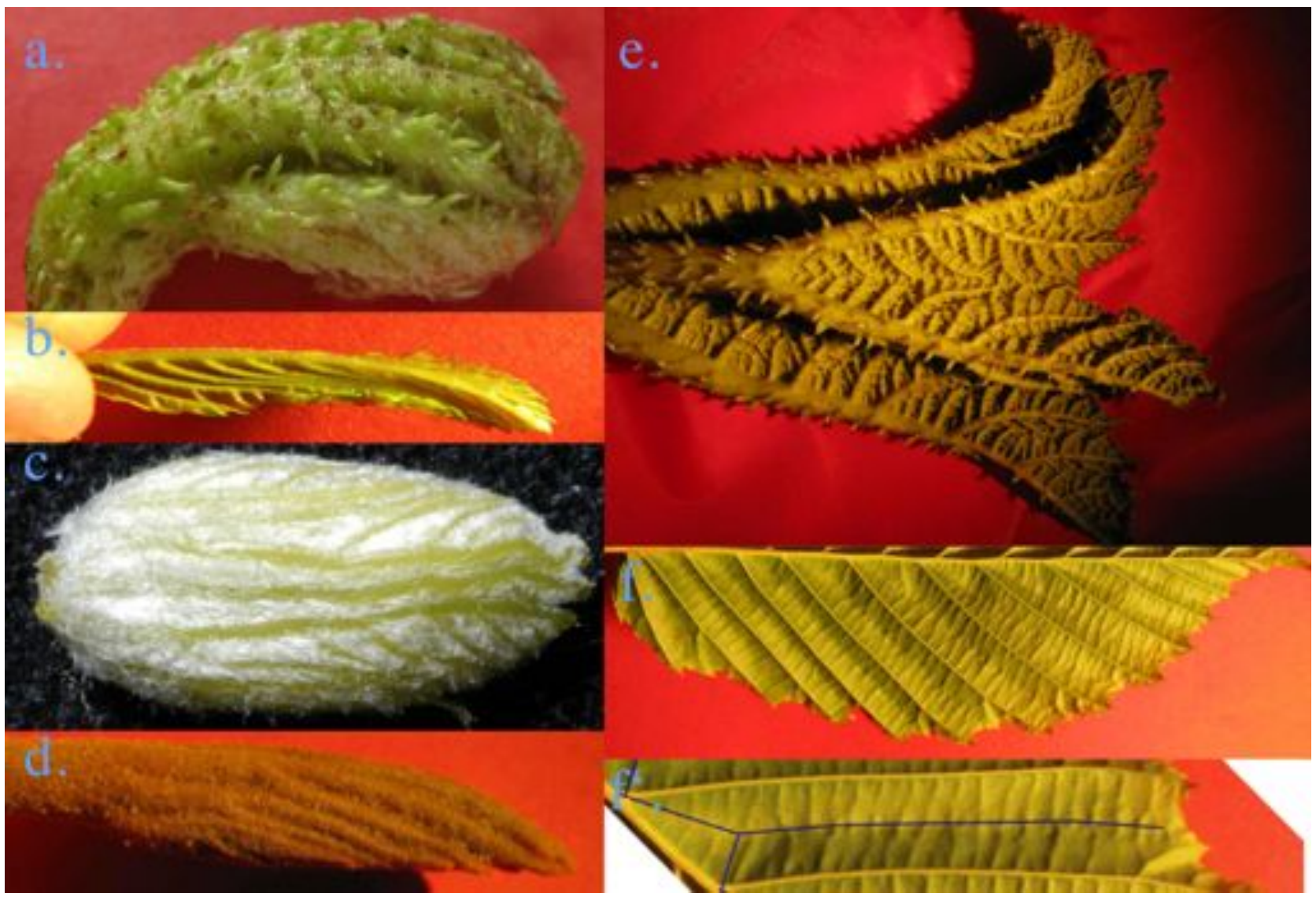

Figure 8: Veins are on the outside of the folded leaf. a. *Gunnera manicata*. b. *Carpinus betulus*. c. *Acer pseudoplatanus*. d. *Tetrapanax papyrifer*. e. As the antifold folds a vein on its neighbouring vein, and the lamina is locally flat, the antifold is situated at the bisectrix of two veins. f. f'. For the same reason, the antifold, which is situated between three veins, is constituted by piece of bissectrix and follows the medial axis of the veins.

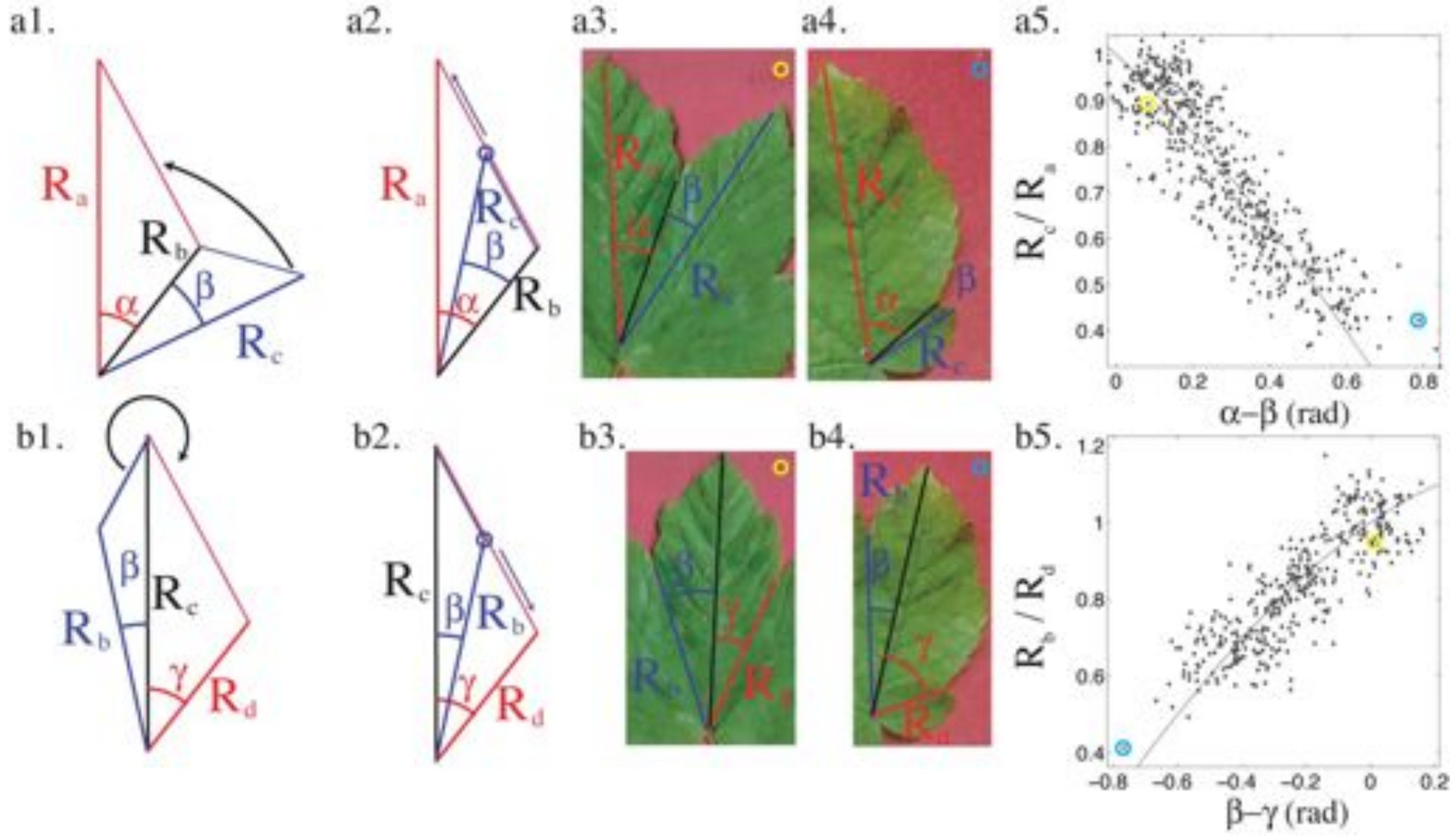

Figure 9: Geometric relationships between two successive lobes and sinuses coming from the Kirigami property. a : Two consecutive primary lobes have veins of lengths $R_a$ and $R_c$. They are respectively making an angle $\alpha$ and $\beta$ with the anti-vein between them, of length $R_b$. b : The vein of length $R_c$ is surrounded by two anti-veins of lengths $R_b$ and $R_d$. These are respectively making an angle $\beta$ and $\gamma$ with the vein. Quantitative relationship on two successive lobes and sinuses coming from the Kirigami property, for 121 *Acer pseudoplatanus* (sycamore) leaves. a5 : Length ratio ($R_a/R_c$) of two consecutive main veins in function of the difference ($\alpha$ - $\beta$) between the angles there are making with the anti-vein. The anti-vein always runs closer to the smaller lobe (a4). When lobes have equal length the anti-vein is then a bisecting line ($\alpha = \beta$, a3). b5 : Length ratio ($R_b/R_d$) of two consecutive main anti-veins in function of the difference ($\beta - \gamma$) between the angles there are making with the vein. The vein always runs closer to the smaller anti-vein (b4). When anti-veins have equal length the vein is then a bisecting line ($\beta = \gamma$, b3).

### *2.4 - Quantitative perimeter symmetry*

#### 2.4.1 – primary folds

Different kind of folding will have different consequences on the leaf shape. The simplest leaf folding, called « radial folding », with folds radiating from the same point, corresponds to the easiest way of folding a sheet of paper. To represent the limitation of the lamina form the enclosing volume (filling law), these folds are cut along a plane. This adaxial plane determines the folded margin of the leaf. After unfolding, the geometry is very simple (Figure 7c).

To find the formula, use the triangle delimitated by $R_a$ and $R_c$ (Figure 9a2) and the perpendicular to the leaf lamina. On this last line, and using the angle p at the main lobe, we can write the relation which link the length of a triangle edges and its angles from $R_a$ and $R_c$. Equalling both gives:

$R_a/R_c=\sin(p+\alpha-\beta)/\sin(p)$.

In the figure 9, we used the average of all the p from all the group of two consecutive lobes to compare the formula with the measurements.

The fact that on both side of a fold, the leaf margin relays on the same line, means, for two consecutive lobes, that the longest is the one whose vein makes the biggest angle with the separating anti-vein fold (Figure 9a). Similarly, for two sinuses around a middle vein, the smaller of two sinuses makes a bigger angle with the middle vein (Figure 9b). This qualitative rule explains why all « radially folded » leaves, namely the palmate leaves, have a similar set of leaves shape, in particular their successively smaller lateral lobes, independently of their phylogenetic origin

(Figure 10 a-c).

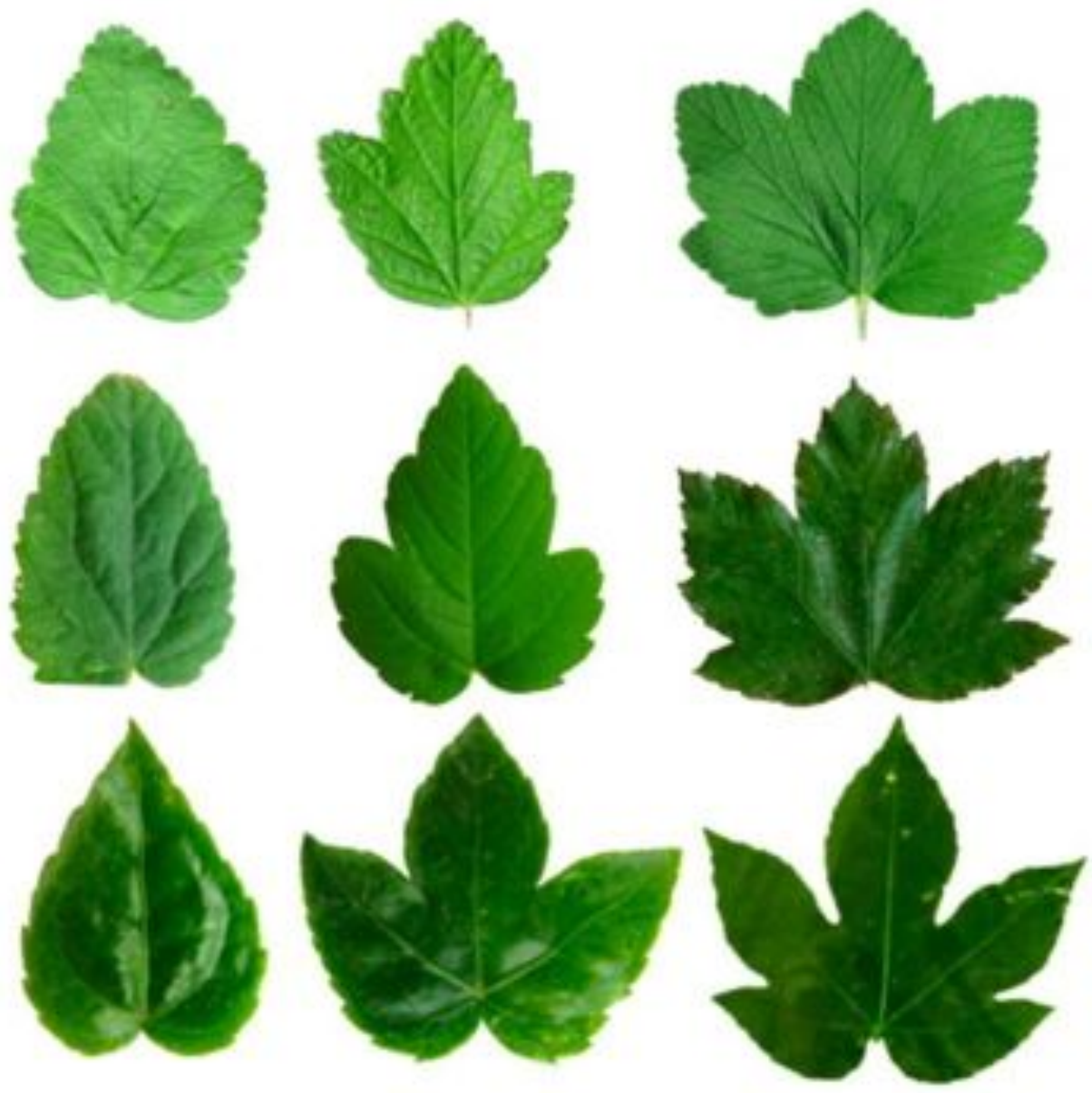

Figure 10: Fold-and-cut leaves from different phylogenetic origins have the same set of leaves shape. a-a": *Acer pseudoplatanus* leaves, *Sapindales*. b-b": *Ribes Nigrum* leaves, *Saxifragales*. c-c": *Fatsia japonica* leaves, *Apiales*.

2.4.2– secondary folds

The second step of complexity in folding is to create secondary folding along a previous fold. After cutting it will create secondary lobes at the margin of the unfolded sheet of paper (Figure 7f). The different dimensions of these secondary lobes will be linked quantitatively. For instance, angle of opening of a sinus can be predicted by using the angle of the precedent sinus and the angle between the folds (Figure 11 a-b).

Using the Figure 11a, we can predict the opening angle $\psi_2$ using the precedent opening angle along the lobe $\psi_1$ , and the angles $\alpha_1$, $\beta_2$, $\gamma_2$ between veins and antifolds. For this purpose, we need to find the value of the angle at the stage where the leaf was folded. As for numerical folding,we have used the angles $\beta'_2$, $\gamma'_2$, which are the nearest angles from $\beta_2$ and $\gamma_2$, which follow the relation of Kobayashi to be foldable in a plane.

We consider the angles of ABCD of the figure 11b:

$$\alpha_1 +(\beta'_2 -\gamma'_2)+(\pi-\psi_1/2)+(\pi-(\pi-\psi_2/2))=2\pi.$$

We can rewrite it:

$$\alpha_1 +\beta'_2-\gamma'_2-\psi_1/2-\pi+\psi_2/2=0.$$

Thus :

$$\psi_2 =2\pi+ \psi_1 -2\alpha_1 -2\beta'_2 +2\gamma'_2 .$$

To test this prediction, we have measured the angles shown on Figure 11a on five leaves of *Tetrapanax papyriferus*. We have calculated the angle of refolding by taking the nearest angle obeying the theorem of Kobayashi (see Annex). We find that the predictions follow the measurements but always under it (Fig. 11c). It is probably because we considered only the curvature due to the fold. As we see on Fig 8d, the vein is curved between the folds too. So we underestimate the curvature of the central vein, overestimate $\alpha$ and then underestimate $\psi$.

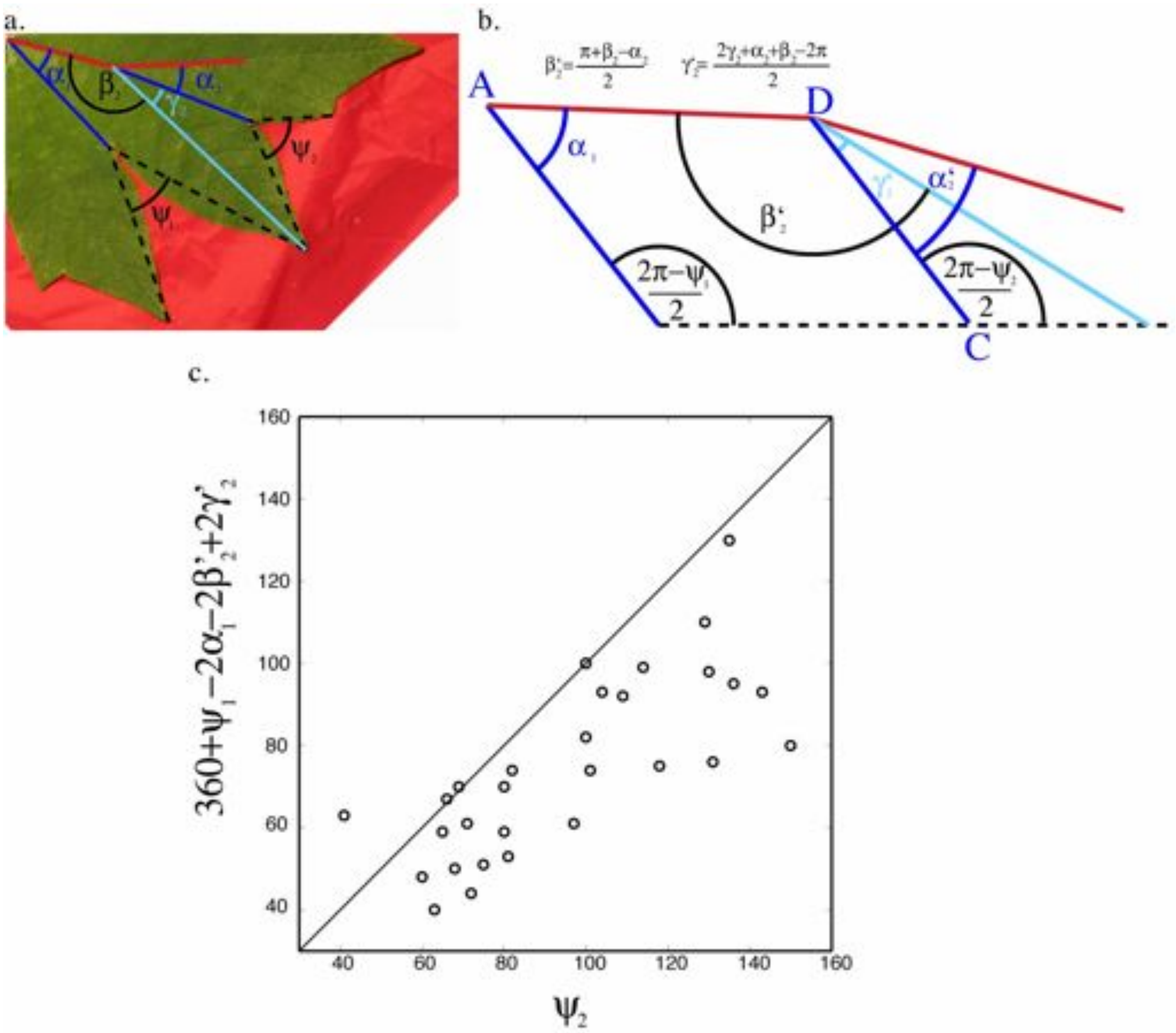

Figure 11. a: A lobe of a *Tetrapanax papyriferus* leaf. The red line corresponds to the main vein. The cyan line corresponds to the secondary vein. The blue line corresponds to the antifold. Notation for the measured angle. $\alpha_1$, $\varphi_1$, $\alpha_2$, $\beta_2$, $\gamma_2$ and $\varphi_2$. b: The same lobe once folded. $\beta'_2$ and $\gamma'_2$ are the nearest angle from $\beta_2$, $\gamma_2$ that you can fold in a plane. It is why we use them to make the prediction. c: Data obtained by measurement on five leaves of *Tetrapanax papyriferus*. We draw in abscissa $\psi_2$ and in ordinate the prediction made by assuming that all the lobe refolds along the same line.

These numerical constrains have a simple consequence on leaf shape: Opening angles $\psi$ of the sinuses increases along each lobe (Fig. 12b-c). If we take into account the folded phase of development, the origin of this observation becomes clear. The key remark is that secondary folds bend the central vein (Fig. 12a, $\beta$ =166° ± 5° - average on 30 folds). Measurements show that angle $\alpha$ between the vein and the antifold does not depend on its place along the main fold, indicating an identical lobe development (28.1° ± 5° for the first angle along the lobe (22 folds), 26° ± 6° for the second one (22 folds) and 27.4° ± 5° for the third (8 folds)). Then, because of the curvature of the main vein which is due to these secondary folds, the angle $\phi$, between the antifold and the cut plane, decreases along the lobe (Fig. 11c). As the angle $\phi$ is smaller along the lobe, the angle of sinus opening $\varphi$, which is equal to $2\pi-2\phi$, becomes larger along the successive lobes (Fig. 12c).

All the secondary lobes have a different orientation toward the adaxial plane. Rather than having an identical shape, as would derive from a simple reiteration process, they adapt their shape to this local geometrical environment to obey the filling law.

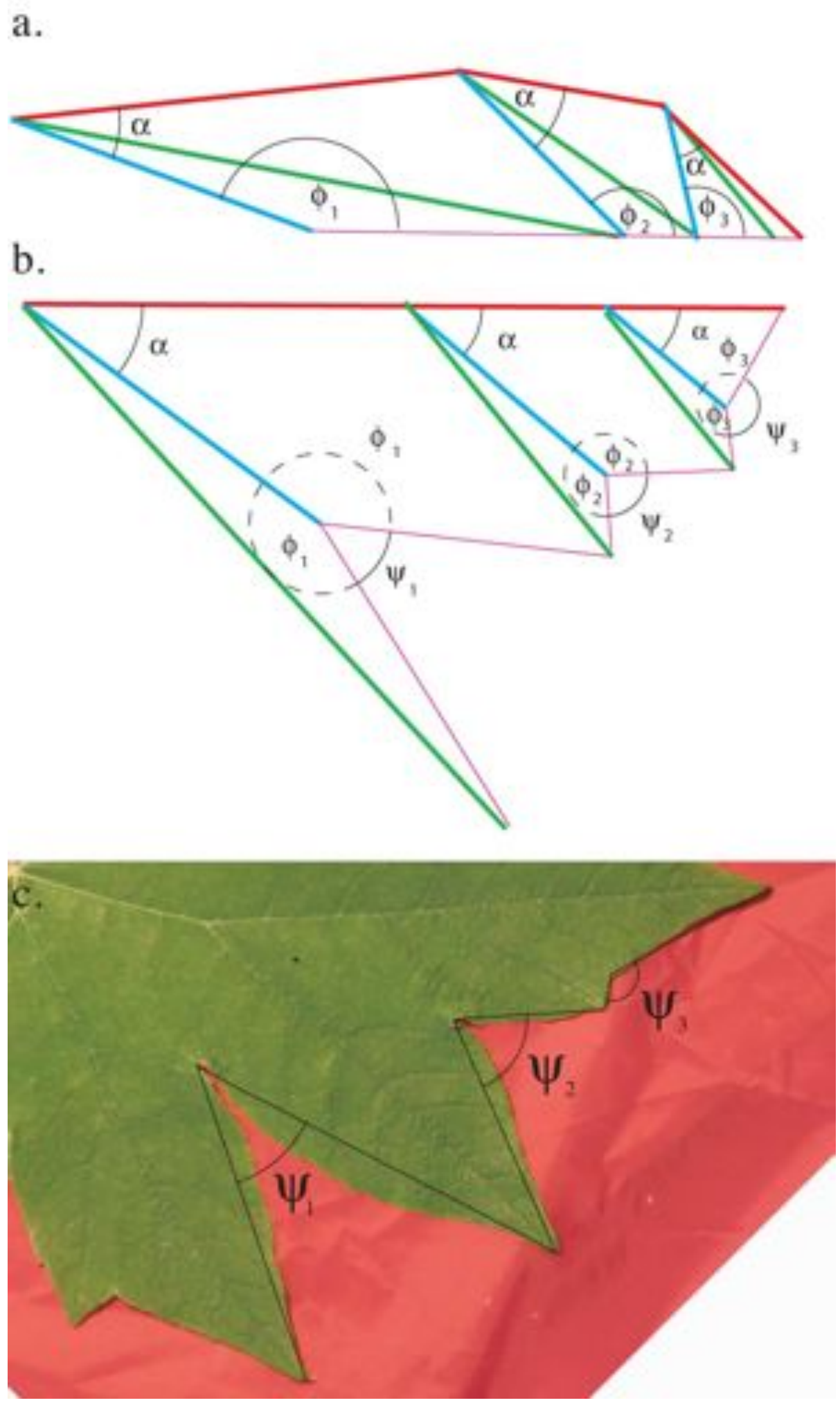

Figure 12 : Consequence of secondary folds. a. Main vein (red) is bent by each secondary folds (green) and antifold (cyan). The angle between antifold and vein does not depend on its place along the vein. For this reason, the angle between the antifold and the cut decreases along the vein. b. For this reason the opening angle of the sinus increases along the lobe. c. Three successive sinuses along a lobe of *Tetrapanax papyrifer.* The opening angle increases : $\psi_1<\psi_2<\psi_3$.

### *2.5 - Non planar folding & Adaxial limitation*

All the precedent folding can be done with a flat sheet of paper (null Gaussian curvature). It is a very restrictive condition. For instance two folds originating from the same point cannot rejoin somewhere else. Then each fold originating from one place of the leaf, as primary folds from the petiole or secondary folds form a vein, can only propagate until they touch the leaf margin, and create either a sinus or a lobe. But some leaves are not flat when they grow (as the antifold of Figure 8f'), nor once grown, as for *Morus platanifolium*. The external shape of a *Morus platanifolium* leaf is as simple as in the "radial folding" case: the abaxial part correspond to the protecting veins, and the margin lays on an adaxial plane delimited by a stipula. But the folds (Figure 13a) can be curved and thus have gaussian curvature, which changes strongly the leaf properties. If the folds are no more straight lines, two folds originating from the same point can reconnect somewhere else, and then cancel each other before touching the edge (Figure 13b1-b2). For this reason, they do not alter the margin. These curved folds enable new kind of folding such as inverted folds: in this case, a fold along a vein, inverts just before the margin and becomes an antifold (Figure 13c1-c2-c3-c4).

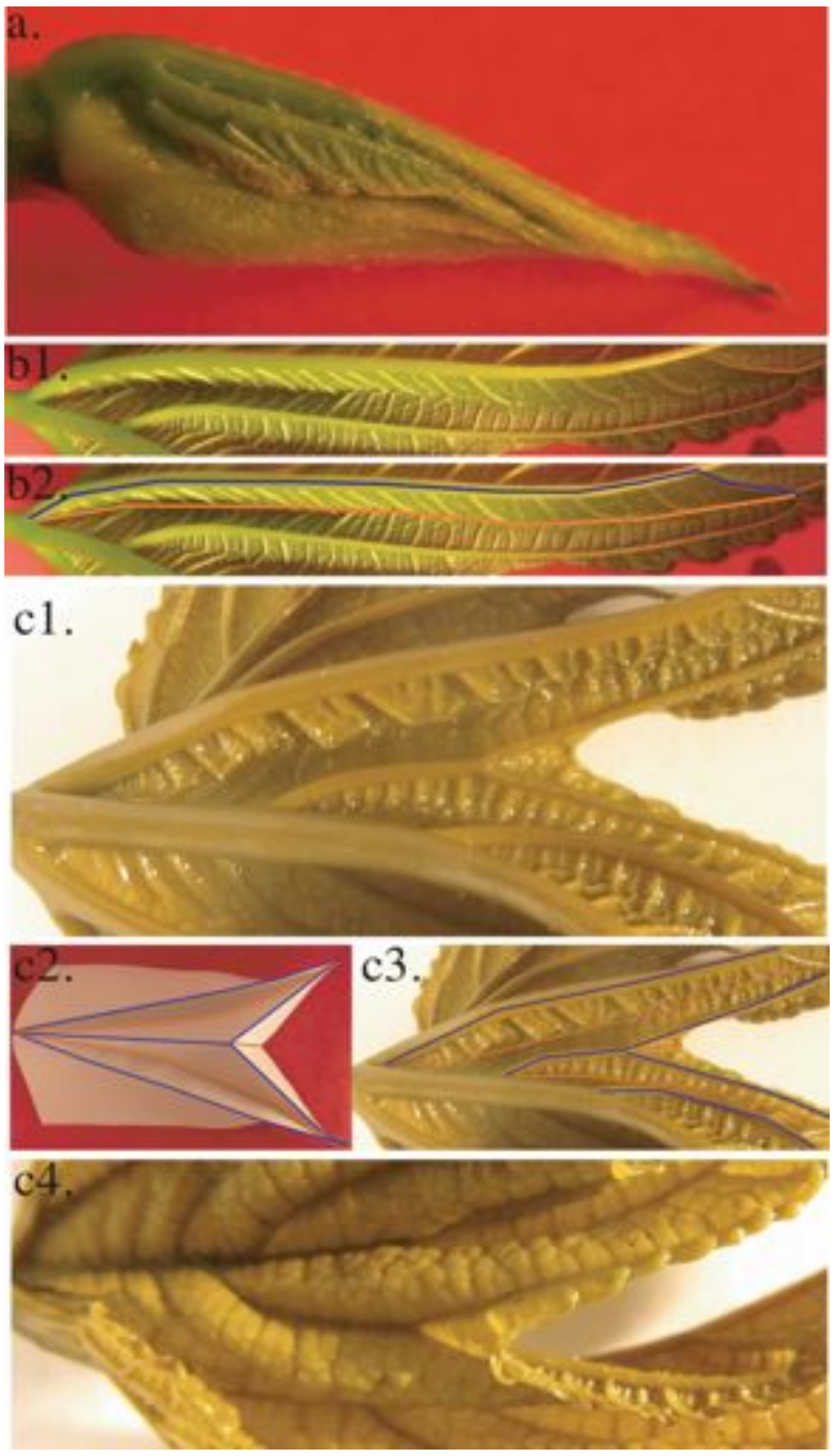

Figure 13. a : A *Morus platanifolium* bud. The leaf is folded with its margin on the smaller bud envelope. b1 : A *Morus platanifolium* leaf fold, which ends before the margin of the leaf without creating sinus. b2 : Sketch of the folds network. (blue line correspond to the folds and orange one to the antifolds). c1 : A *Morus platanifolium* leaf fold (along a vein) invert itself in an antifold just before the leaf margin (abaxial view). c2 : Simplified fold network using a sheet of paper. c3 : Fold network of c1. c4 : Adaxial view of the same inverted fold.

If two curved folds cancelling each other do not affect directly the margin, they have an indirect effect on the shape: antifolds can be axis of symmetries of lobes (Figure 14b-b'), and veins can be axes of symmetry of sinus (Figure 14c-c'). The symmetry rules appears to be the inverse of classical Kirigami (Figure 14d), but they can be recovered, considering some topological rules like fold+antifold+fold =fold, so that the antifold in between two folds becomes a symmetry axis of a lobe (and not a sinus, Figure 14b-b'), or similarly antifold +fold +antifold =antifold, and a fold becomes the symmetry axis of a sinus (and not a lobe, Figure 14c-c').

By taking into account these mute folds, we can again refold the leaf on the simple volume it occupies in the bud. (Figure 15)

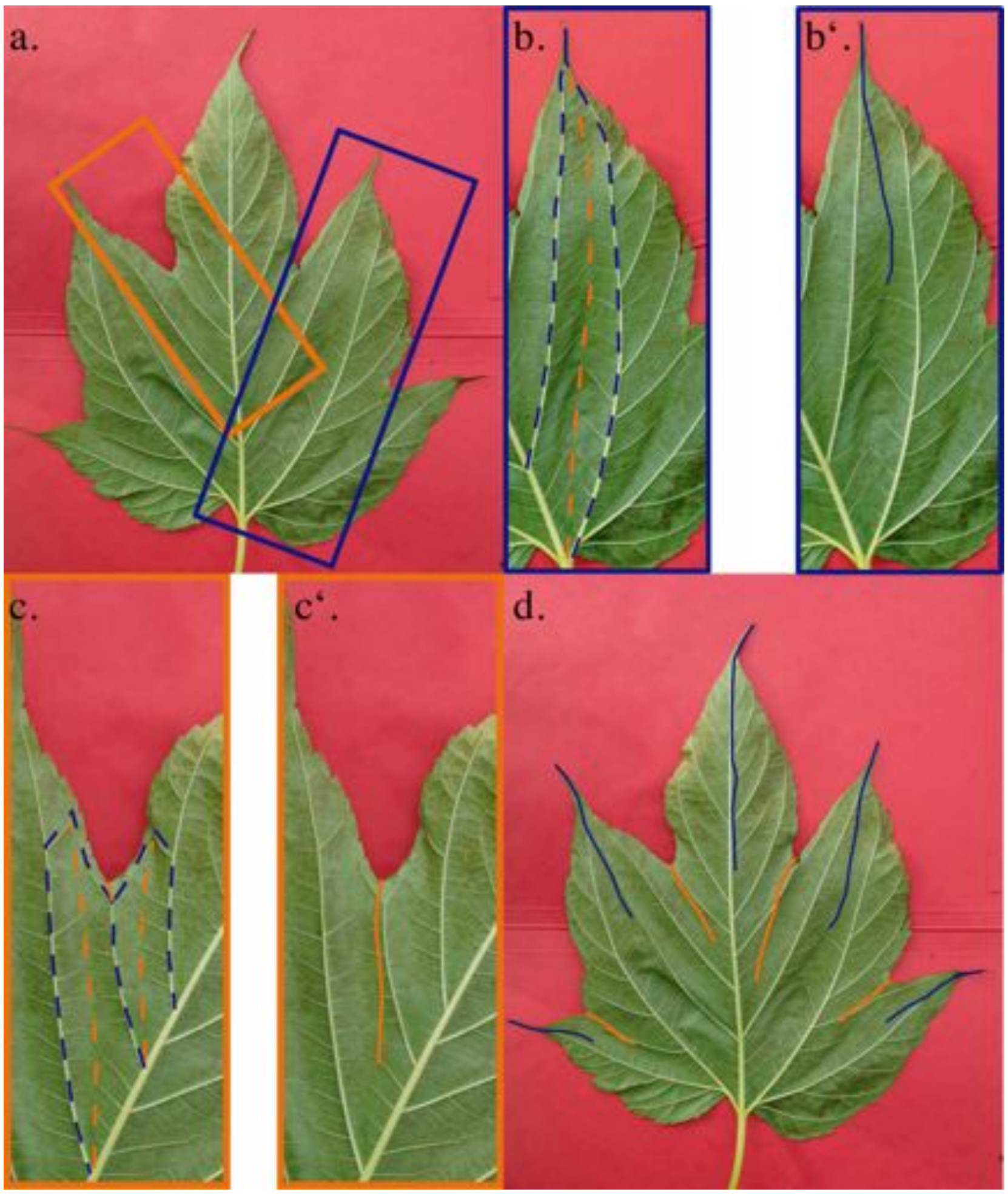

Figure. 14 : Symetry rules are inverted in non planar leaves. a : A *Morus platanifolium* leaf. b: The most basal fold (dashed blue line) originates at the same point than the antifold (orange dashed line), the both join at the lobe tip without creating either another lobe either another sinus. b'. The medial axis of the lobe (blue line) follows the antifold and not the fold as in the case of *Acer pseudoplatanus* (Figure 5c). c: The folds (dashed blue line) and the antifolds (orange dashed line) create a complicated network, which corresponds to an inverted fold. c' : The symmetric of the medial axis of the sinus (orange line) follows mainly the vein and not the antifold as in the case of *Acer pseudoplatanus* (illustration 5b). d : A *Morus platanifolium* leaf. Symetry rules are inverted : The symmetric of the medial axis of the sinus (orange line) often correspond to veins, and the medial axis of the lobe (blue line) often correspond to antifolds.

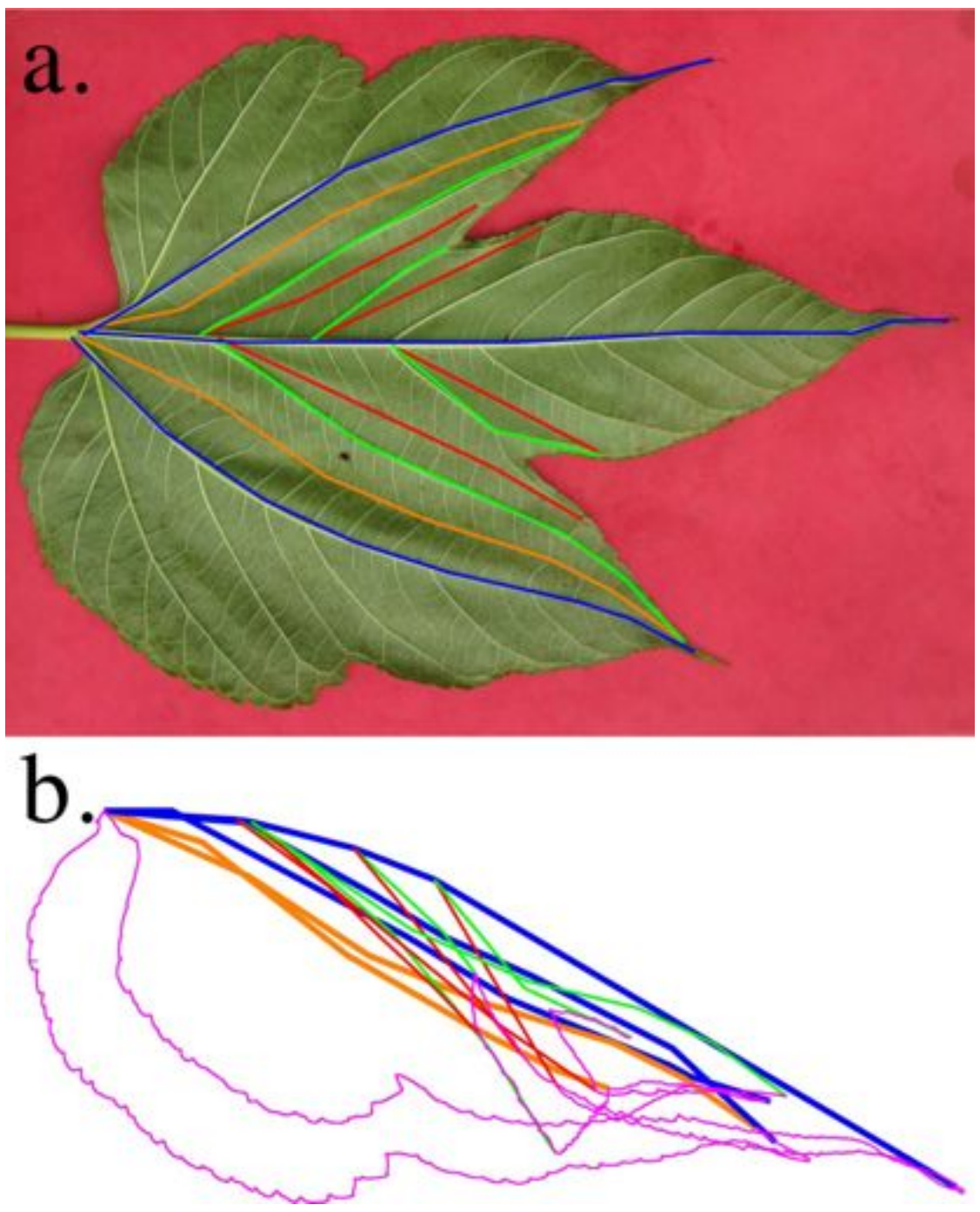

Figure 15. a: a mature leaf of *Morus platanifolium*. b: same leaf numerically folded back. The leaf contour is pink. Primary and secondary veins are respectively blue and red. Primary and secondary anti-veins are yellow and green respectively. Only veins ending at peaks are represented and stand for anticlinal folds along segments linking two consecutive branching points or a branching point to a peak. Synclinal folds run along segments (anti-veins) linking a sinus to the branching point of the two surrounding veins. The thickness of the leaf is not taken into account and the leaf is folded back onto a plane, holding the angles to the best (see Annex).

### *2.6 - Adaxial limitation & self limitation.*

Another particular case is exemplified by *Ficus cariaca*, which is close to *Morus platanifolium* in the phylogeny. The volume occupied by the leaf in the bud is still a quarter of an ellipsoid, delimited by the previous and next stipula (Fig. 16a), also with curved folds. However a small detail in the folding leads to a drastic change in the leaf shape. As in other folded leaves, the central fold develops its lamina partly behind the lateral one, partly in front of the stipule of the smaller bud. Apparently the whole leaf margin lay on the stipule plane (Fig. 16b). But if the central lobe of the leaf is moved slightly away from the lateral lobes, the edge of the lobe shows to run along the vein of the lateral lobe and does not extend up to the stipule plane, except at its end where there is no longer any secondary veins (Fig. 16b'). This way of being folded results in the spoon like shape of lobes. The base of a lobe is thin because it is limited by the lateral veins (Fig. 17). The end of the lobe is large because it is limited by the bud in front and no more by the lateral vein (Fig. 16c).

This relationship between the base of a lobe and its lateral vein come clearly by regulation, not by chance. This can be checked by the fact that the contact is always perfect while the global shape of the leaf is varying strongly by the size and number of lobes. Along the stipula, the ratio between the length of the edge of the lobe tip and the rest of the margin can vary from 0.4 to 0.9 (Figure 18a-b). However the central lobe edge always fits the lateral veins. Is it the vein, which fit to the edge or the contrary?

If we refold the leaf we see the ends of all lobes almost align, whereas the edge of lobe's base stays along the next folded vein (in blue, Fig. 16g-h). The numerical folding of the mature leaf (Fig 18b-b') brought the lobe edges close to a plane, despite latter expansive growth somewhat reducing the fit to a plane compared to the young folded leaf in Fig 16b.

The geometry of the packing in the bud left its imprint on leaf shape. The fig tree leaf shows that the filling law is very strong: it remains true for a curved limitation and a limitation by two distinct objects.

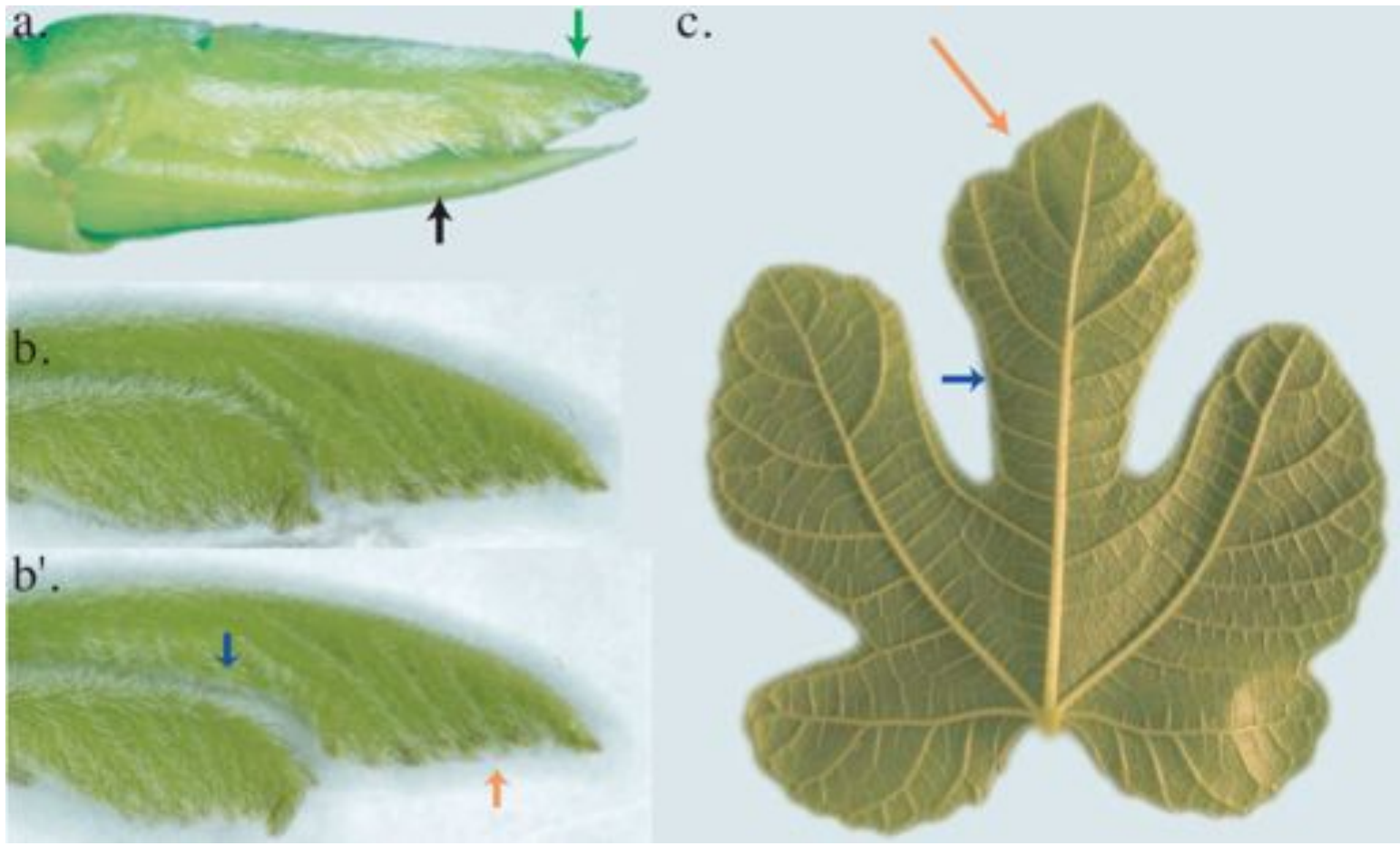

Figure 16. a: A *Ficus Cariaca* bud. The leaf (green arrow) is limited by another smaller bud in front of it (black arrow). b: A leaf of Ficus cariaca alone.  b’: the same leaf with the central lobe slightly moved. It reveals that its lower border lies on the lateral lobe (blue arrow), except at the end where it remains along the stipula (orange arrow). c: A mature *Ficus Cariaca* leaf, showing the two corresponding parts, the side vein limited part (blue arrow) and the stipula limited part (orange arrow).

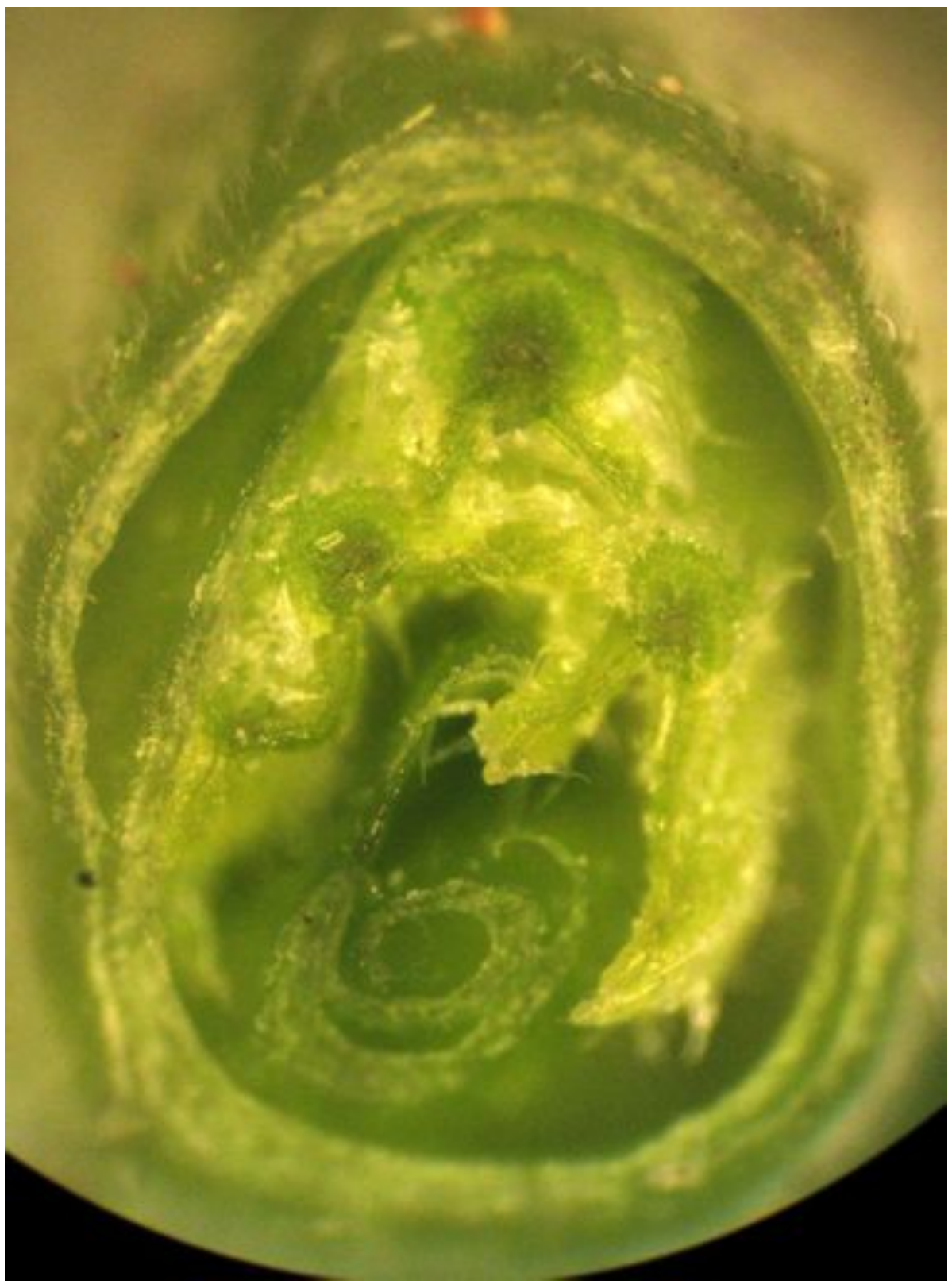

Figure 17: A cut of a *Ficus Cariaca* bud. The lamina folded around the central vein ends near the lateral veins (arrows), while the  lamina of the lateral lobes ends at the lower stipula.

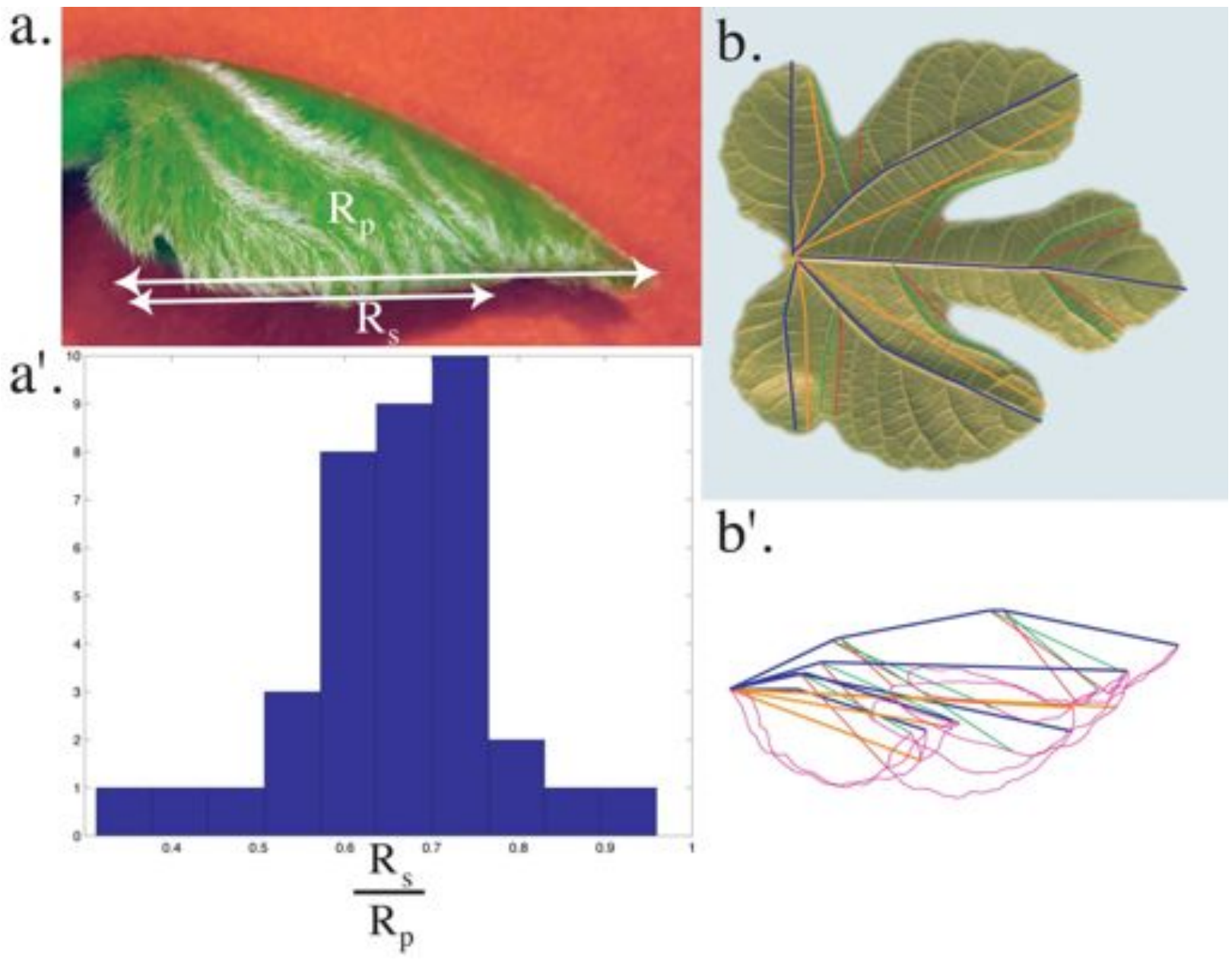

Figure 18. a: $R_p$ is the length of the entire edge of the leaf. Rs is the length of the edge until the beginning of the central lobe. a': Histogram of the ratio $R_p/R_s$ , showing large variations. b: A mature leaf o*f Ficus Cariaca.* Color corresponds to the different lobe. Blue for principal fold. Orange for main antifolds or valley folds. Green for secondary folds. Red for secondary sinus. b': The same leaf once folded numerically (see Annex).

### *2.7 - Abaxial limitation & transverse cut*

We can imagine that the cut delimiting the leaf border is no more orientated toward the adaxial plane but toward the abaxial surface. The folds are always along the veins, which are continguous with the abaxial envelope. In the previous (common) case, it is the adaxial plane that cut the folds transversally (Fig. 19a). In the case we consider here, it is also the abaxial limitation that cut the folds, then tangentially (Fig. 19b). For purely geometric reason, it changes considerably the geometry of the leaf border. In the case of transverse folding, folds are axes of symmetry of the edge of this filling surface (Fig. 19a-a'). But if the cut direction is tangent to the fold, the fold is no more an axe of symmetry of this margin (Fig. 19b-b').

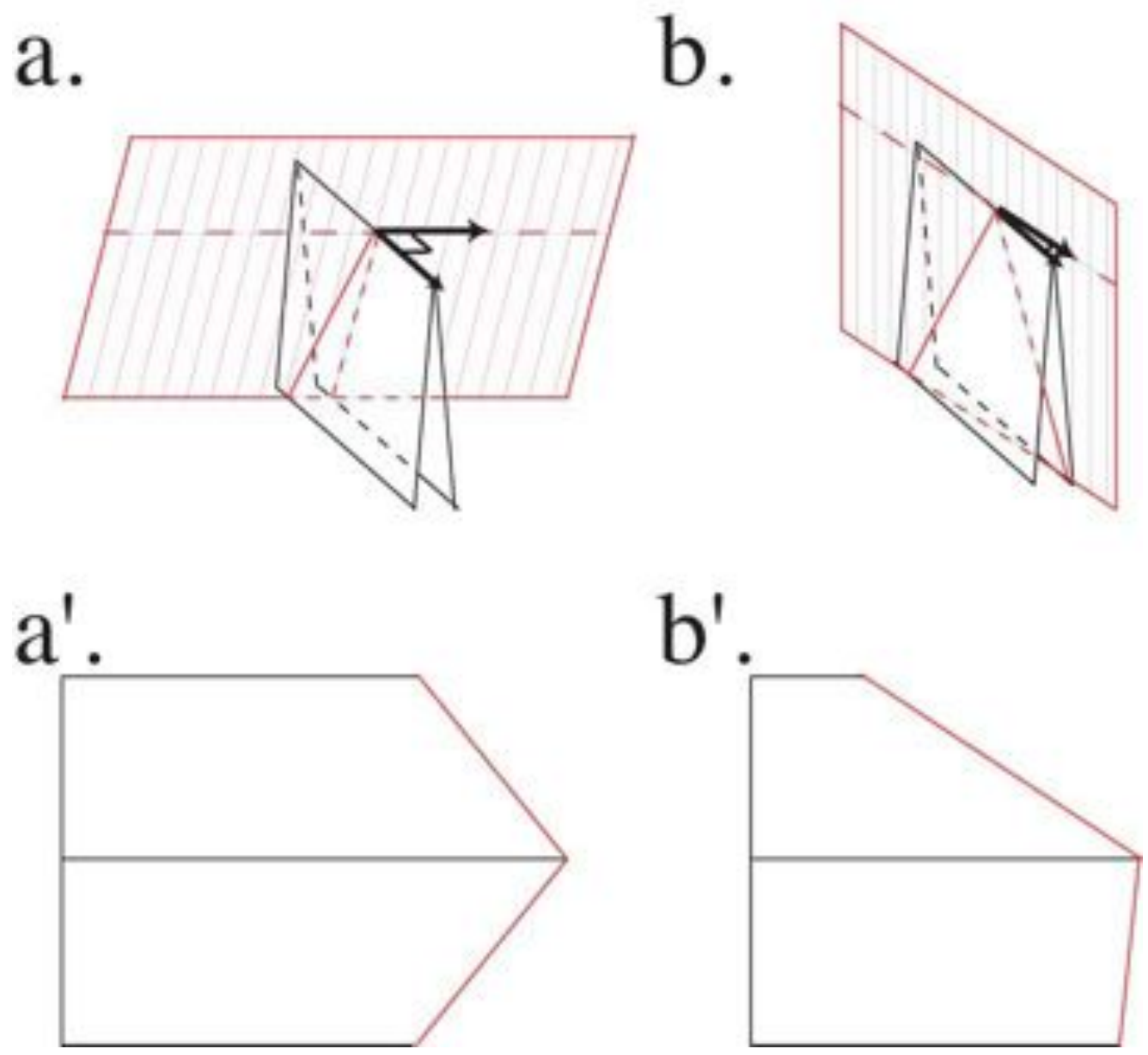

Figure 19. a : Transverse cut of a fold. a' : Same fold than a once unfolded. The fold is axis of symmetry of the edge. b : Tangent cut of a fold. b' : Same fold than b once unfolded. The fold is no more axis of symmetry of the edge

It is particularly interesting to notice this transition between adaxial and abaxial limitation, and thus transverse and tangent cut, can be obtained inside the same species but just between two cultivar, as for beach, between the cultivar *Rohan obelix* (var.) and normal beach. In the case of *Rohan obelix*, the cutting plane is adaxial, corresponding to the back of a younger leaf (Figure 20a). Limitation is transverse to the fold, as common otherwise. As a consequence, folds are axes of symmetry of the edge (Figure 20b-c).

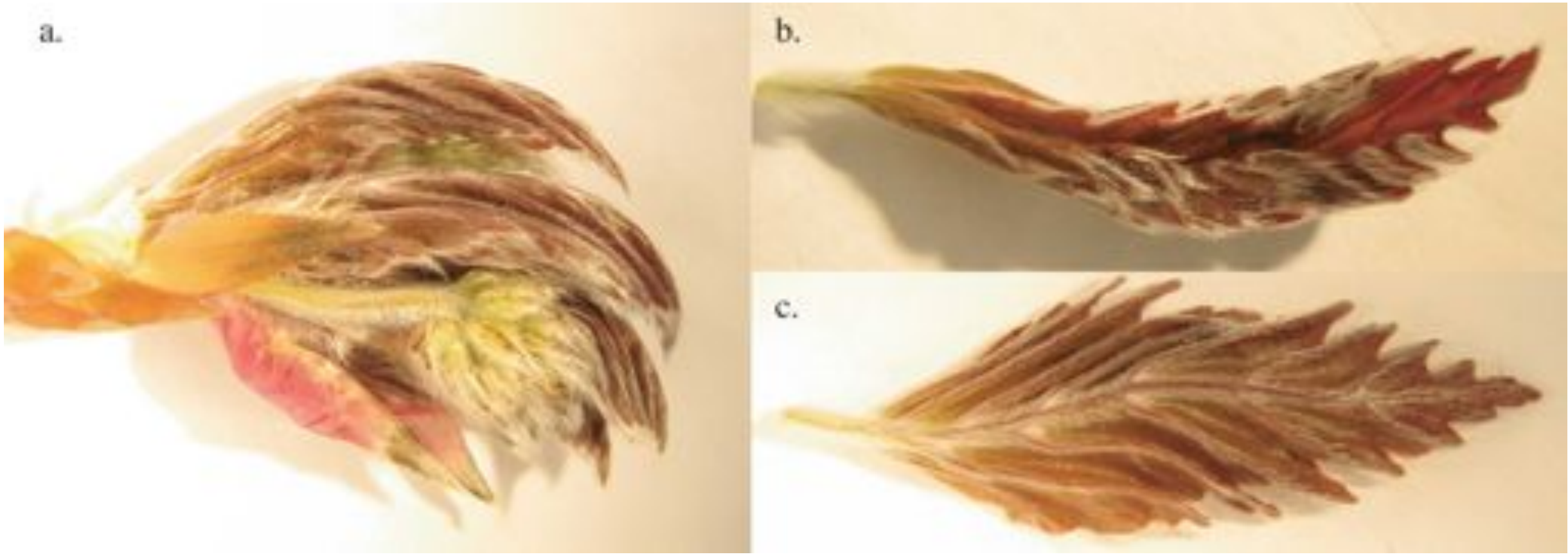

Figure 20. a : A *Fagus sylvatica (Rohan obelix var.*) bud. The leaf margin is folded on the back of the previous leaf, transversaly to the fold. b: Front (adaxail) view of a *Fagus sylvatica (Rohan obelix var.*) leaf and c: back (abaxial) view. Folds and antifolds are axis of symmetry of the margin.

In case of normal *Fagus sylvatica*, the limitation of each leaf is on its abaxial side, contrary to all the other species of kirigami leaves (Fig. 21b). The limitation corresponds to the envelope of the bud, which is tangent to the folds on the sides (Fig. 21a-b). As a consequence, folds are no more axes of symmetry of the margin (Fig. 22c). Even in this configuration the surface of the leaf fills the volume at its disposal (Fig. 21c).

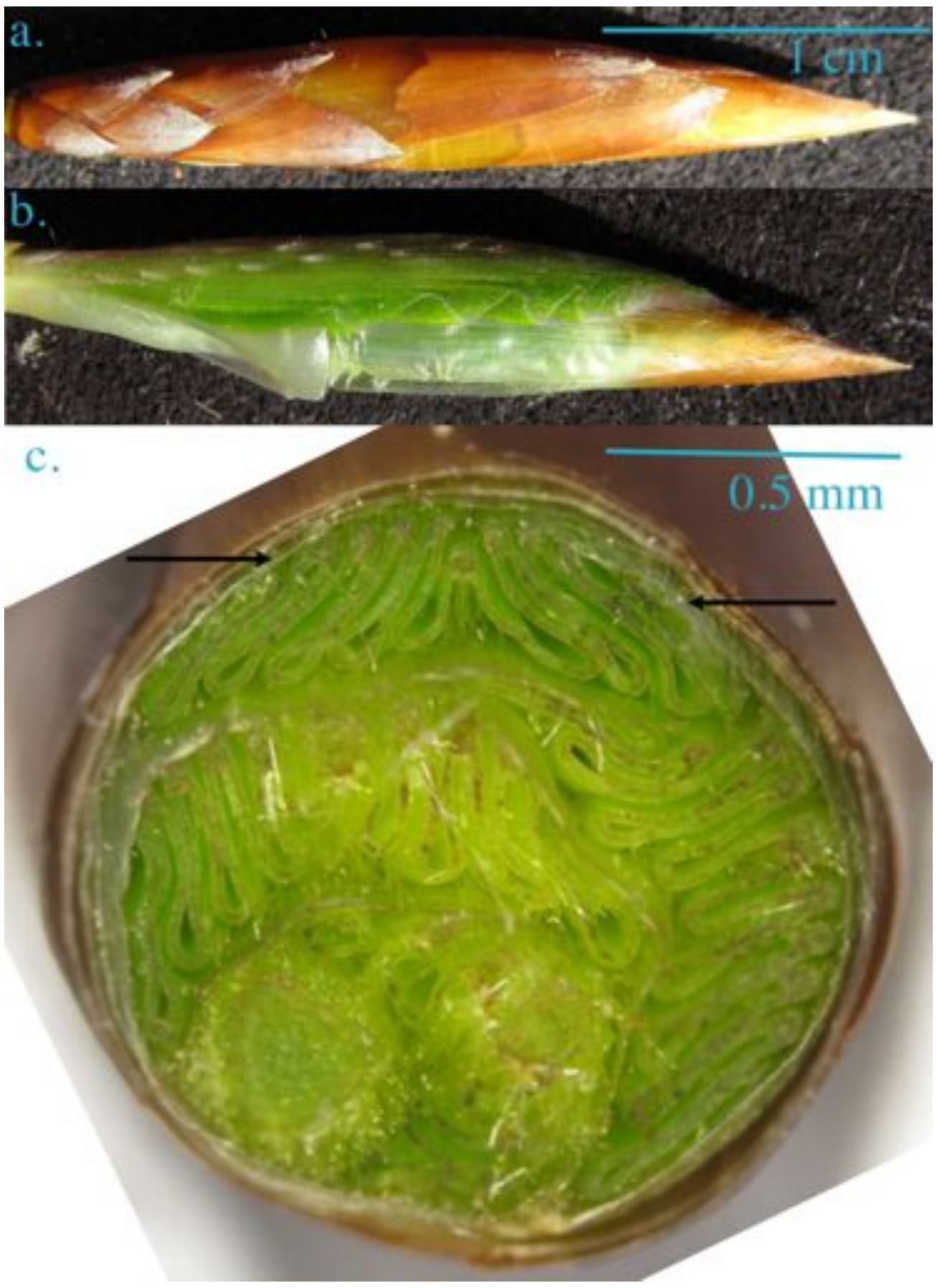

Figure 21. a: A bud of *Fagus sylvatica* b: The same bud whithout its upper outer-shell. The fold of the leaves are not axes of symmetry anymore. c: A transversal cut of a bud. The folded leaf margin lays on the abaxial envelope. (black arrows).

A numerical relationship linking the angle of asymmetry of each fold cut, α, and the angle between fold and limiting surface, β, can be derived as detailed in Figure 23d-e: $\cot(\alpha/2) = h \tan(\beta)/e$. To verify that the border of beach leaves follow this relation, we have taken folded beach leaves and measured the angle β between the fold and the contour of the folded leaf (which corresponds to the outer-shell) and the angle α of asymmetry of the cut (Fig. 23a-b-c). These angles follow the predicted law (Fig. 23f). When β is small, the fold is tangent to the limiting surface and the asymmetry α is large. When β is bigger, asymmetry is much smaller. We observe that on real leaves. The first folds along the central vein are tangent to the outer-shell. They are very asymmetric. On the contrary the last one are nearly perpendicular to the outer-shell so the folds are much more symmetric (Fig. 22c). We can also do the same measurements for other beach leaves. We see that all the leaves measurement align on the same curve (Fig. 23g). It means that cutting geometry is conserved among these leaves.

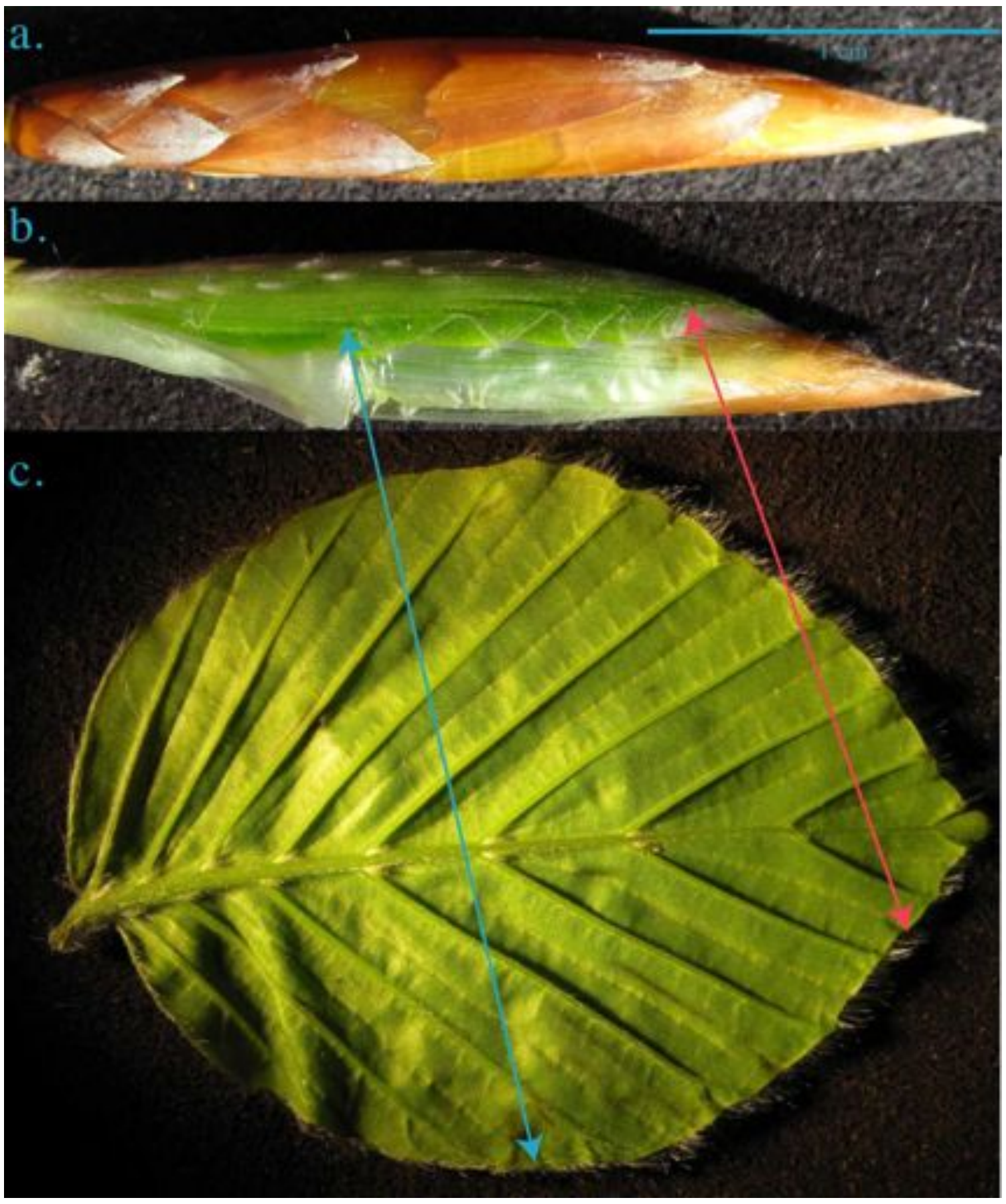

Figure 22. a. A bud of *Fagus sylvatica* whithout its upper outer-shell. The fold of the leaves are not axes of symmetry anymore. b: A mature leaf. First folds along the central vein are not axis of symmetry of the margin (blue array) because they correspond to folds tangent to the outer-shell. Because the folding vein are also one above each other, the antifold is also not at the middle between the two nearby folds. Last folds along the central vein are axis of symmetry of the margin because they correspond to fold which are transverse to the outer-shell (red array), with also veins at the same height and in the middle of the folds.

Tangential cut result in leaves whose antifolds do not correspond always to sinus (Fig. 24). It depends precisely on where the cut is between two consecutive veins. If the antifold cut is closer to the tip of the first vein then the antifold will give a sinus as normal (fig 24a-a'). But if the cut is closer to the tip the second vein, the antifold will end in an unusual small bump between the veins (Fig. 24b-b'). Both types of leaves exist. All these different observations show that beach leaves are subjected to transverse cutting.

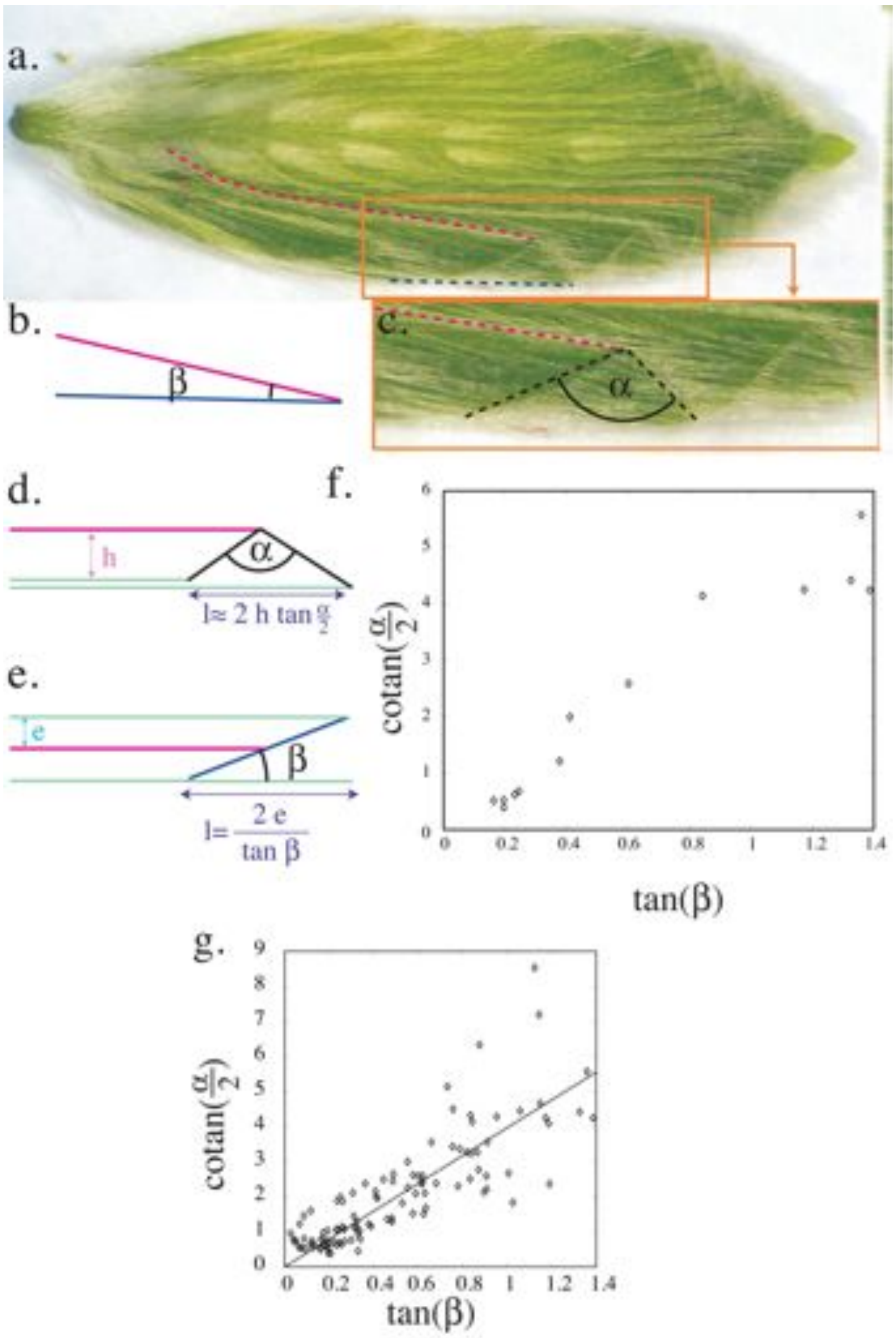

Figure 23. a : Schema of our measurements on the abaxial view of a beach leaf. Magenta dashed line corresponds to mountain fold. Blue dashed line links three consecutive sinus centered around the sinus at the valley fold tip just after the mountain fold. Blue dashed line corresponds to the limiting surface. b : β is the angle between the fit of the blue dashed line (represented by a blue continue line) and the fit of the magenta dashed line (represented by a magenta continue line). c : The angle α (black) corresponds to the asymmetry. d : Schema of a lateral view of a fold. First way to calculate the length l with the angle of asymmetry α. e : Schema of an upper view of a fold. As the tissue has a positive thickness e the two valley folds (red line) do not superimpose with the mountain fold (black line). Second way to calculate the length l. f : cotan(α/2) as a function of tan(β) for the leaf a. g. cotan(α/2) as a function of tan(β) for 8 different leaves.

This example of beach shows that the filling law remains true even in this extreme case where folds are orientated toward the abaxial part and cutting tangentially the leaf border. The change in the leaf shape and cutting of the folded leaf, even between two cultivars, show that this filling law is also very stable while the detail of the cutting geometry can change rapidly.

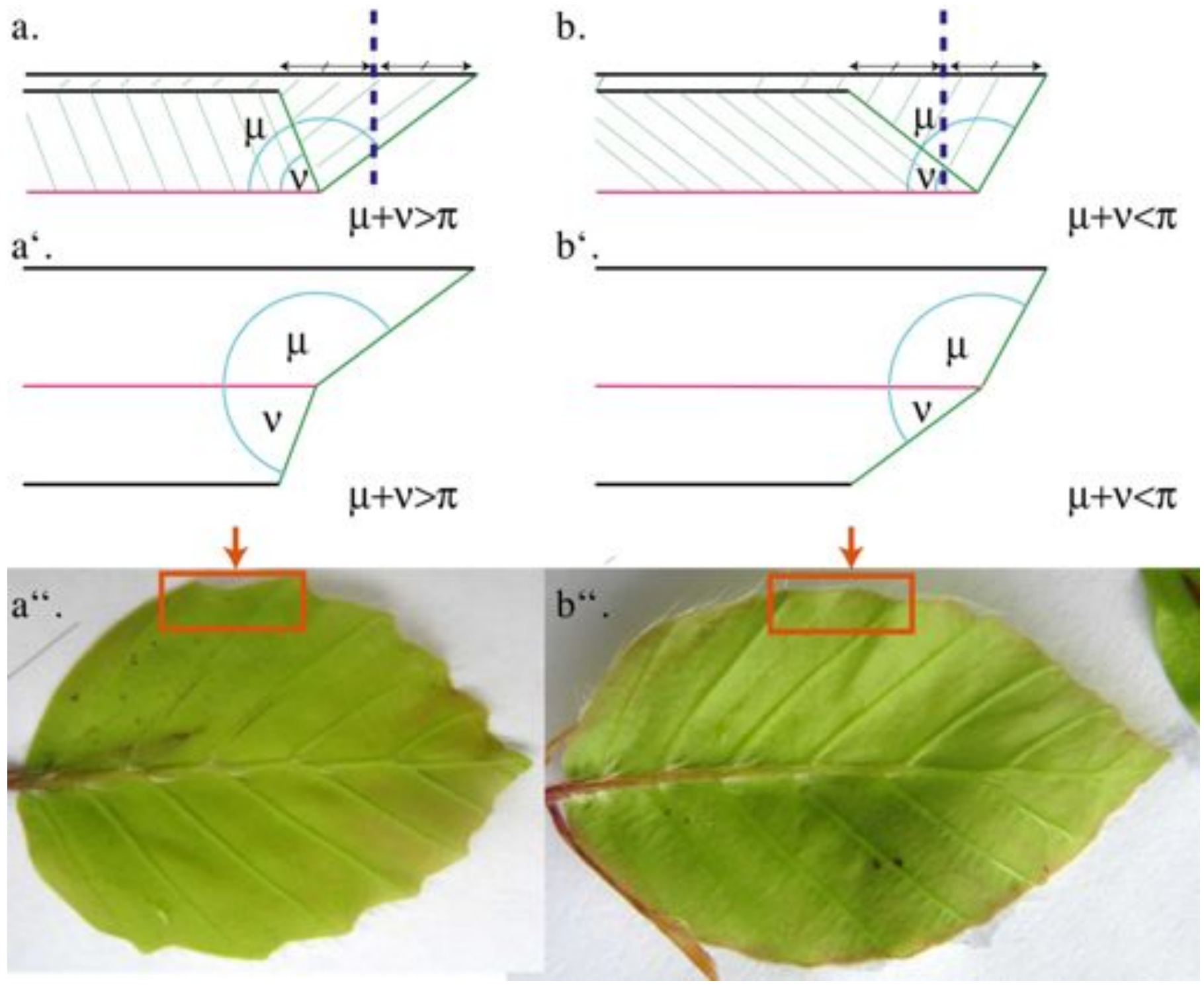

Figure 24. a: The black lines correspond to veins. The red lines correspond to an antifold or "valley" fold. The thick green lines correspond to the leaf margin. If the antifold's end is nearer from the tip of the first vein than from the tip of the second one then $\mu+\nu>\pi$. a': It will give a sinus when it will unfold. a'': Beach leaf whose antifolds become sinus - red arrow. b-b'-b'': If the antifold's end is nearer from the tip of the second vein than from the tip of the first one then $\mu+\nu<\pi$. It will give a beach leaf whose antifolds become lobe - red arrow.

## 3 – Discussion

### *3.1 - Regulation*

Even if all these leaves shapes are very diverse, a common relation links them to their folding. Leaf shape seems to be the consequence of filling the bud with a certain folding, and not the contrary. Without doubt evolution has taken into account this bud packing constrain in the leaf shape.

We can thus ask how this packing constrain is encoded in the leaf development. From the point of view of developmental genetics, lobes are reiteration of primordia on the primordia itself. First they are constituted by an axis, which will give the vein. The lamina emerges symmetrically from the both side of this axis.

If this theory explains easily that veins are axis of symmetry of the lobe margin. But it does not explain why antifolds are axes of symmetry of the sinuses. It wouldn't explain also why the secondary lobe angles are different as in *Tetrapanax papyriferum*. Kirigami rules of folding and cutting seems to be stronger than this rule of symmetry of the margin around a central vein, as suggested by both the *Morus* and *Fagus* cases: because of mute folds, veins are not axes of symmetry in *Morus platanifolium*; because of tangent cut, neither the veins nor the antifolds are axis of symmetry of the *Fagus sylvatica* leaf margin.

We could argue thus argue that the Filling law does not come from a pre-controlled development, but from a global regulation during the development. A possible candidate for integrating the packing constraint is contact regulation (by the constraining volume). If contact regulation is very common at higher scale of vegetal development like at the scale of the tree (an example is the shyness of crown) [Putz et al (1984)], or at the scale of the stem or organ expansion [Coutand and Moulia (2000)], it has never been studied at the scale of leaf development. Recently researchers have shown that mechanic plays a role in primordia development [Hamant et al (2008)]. Contact regulation at the scale of the embryonic leaf would close the gap between the scales. The contact limitation did not appear earlier in developmental biology studies possibly because researcher have studied free flat leaves as the *Arabidopsis thaliana* ones. On the contrary, the leaves studied here are stacked, and are surfaces constrained by each other in a tree dimensional bud. One conclusion is that folded leaves should not be considered as simple surfaces when they develop but as fully three dimensional objects. Conflict for space in the bud and its consequences can't be discarded, and there is a need for contact regulation.

The difference between two cultivars of beech also shows that a very small change (cultivars are still interfertiles), can change the way the leaf are folded in the bud, and have a direct consequence on the leaf shape. In this way the details of the packing geometry seem very variable, while the packing regulation remains unperturbed.

Our observations clarify how the interplay between reiteration and regulation in lobe development could work. The *Tetrapanax papyriferum* secondary lobes show that the fold angles are independent on its place along the main vein, whereas the parameters of the corresponding lobe adapt so that its border meet the enclosing surface. In this way there seems to be a reiteration of identical folds, but the expansion of the lamina, and thus the lobe angle, adapts to the available volume. Fold seems to be the real unit of reiteration rather than the lobes, which shape are then given by the volume limitation.

*3.2 - Evolution*

The fact that leaf grows folded could be evolutionary interesting. The folds are not random (like in expanding poppy petals), but strongly related to the development of the veins. The first interest is to present the veins on the outside. The veins are large and robust, mainly composed of spongy tissue which first role is mechanical (parenchyma), but also provide isolation. When damaged, the phloem can give a quick chemical response. Thus the veins covering the whole exterior of the embryonic leaf provides a good protection, against cold, dryness, and predators (insects). The rest of the bud structure is also aimed at protecting the young leaf.

It suggests also an original evolutive interest for the lobed shape of leaves. If the shape of a leaf without lobe is simpler than a leaf with lobes, once they are folded, on the contrary, the lobed one is much simpler and more compact (compare for instance the figure 7a of a folded sheet of paper, which is complicated, and the figure 7b of a folded lobed sheet of paper). The evolutive interest of the lobes might be indirect. It would be an efficient way to develop the largest leaf surface in a volume with the smallest external one. In temperate climate, it could be a good mean to minimize the loss during winter, and to maximize the photosynthesis as soon as spring comes. The perspective is coherent with the observation that in one specie the folded shape of the leaves (with its margin on a plane, or other fold kind) is much more conserved than the final unfolded leaf shape, which has a great variability (as we can see in figure 9 for instance). It means that much more energy is invested by the plant to regulate the shape of the leaf at this folded stage rather than later. As phyllotaxis [Douady and Couder (1996))], the particular result would just be a consequence of a dynamical packing problem. The shape of the leaf would just be the consequence of the details of the growth of each bud, their particular geometry and growth history.

This is very different from the flowers (petals) regulation where the direct control of the final shape [Rolland-Lagan et al. (2003)] would be important to achieve a good reproduction, the pollinator reacting to the shape (and color) of the petals, while the interest of the leaf is just to produce an irrigated surface for photosynthesis, independently of its shape. The various presence of the Kirigami property along the whole evolutionary tree also show that it is not highly stabilized property, contrary to flower shapes, and also that is based on basic mechanisms that can be switched on and off easily.

The aim of growing folded leaves as a protecting strategy could finally explains the observation of the predominance of palmate leaves in cold-temperate regions [Bailey and Sinott (1916)], where this protection is most needed. Some buds have an acute shape (Ficus cariaca, Fagus sylvatica), which might be a protection against herbivors. Interestingly, these acute buds correspond to most complicate folding (Figure 25). Leaf shape evolution could then be an indirect consequence of an upper pressure on the bud shape. For instance it has been found in tropical rain forest that enrolled monocotyledons buds are less eaten [Grubb *et al* (2008)].

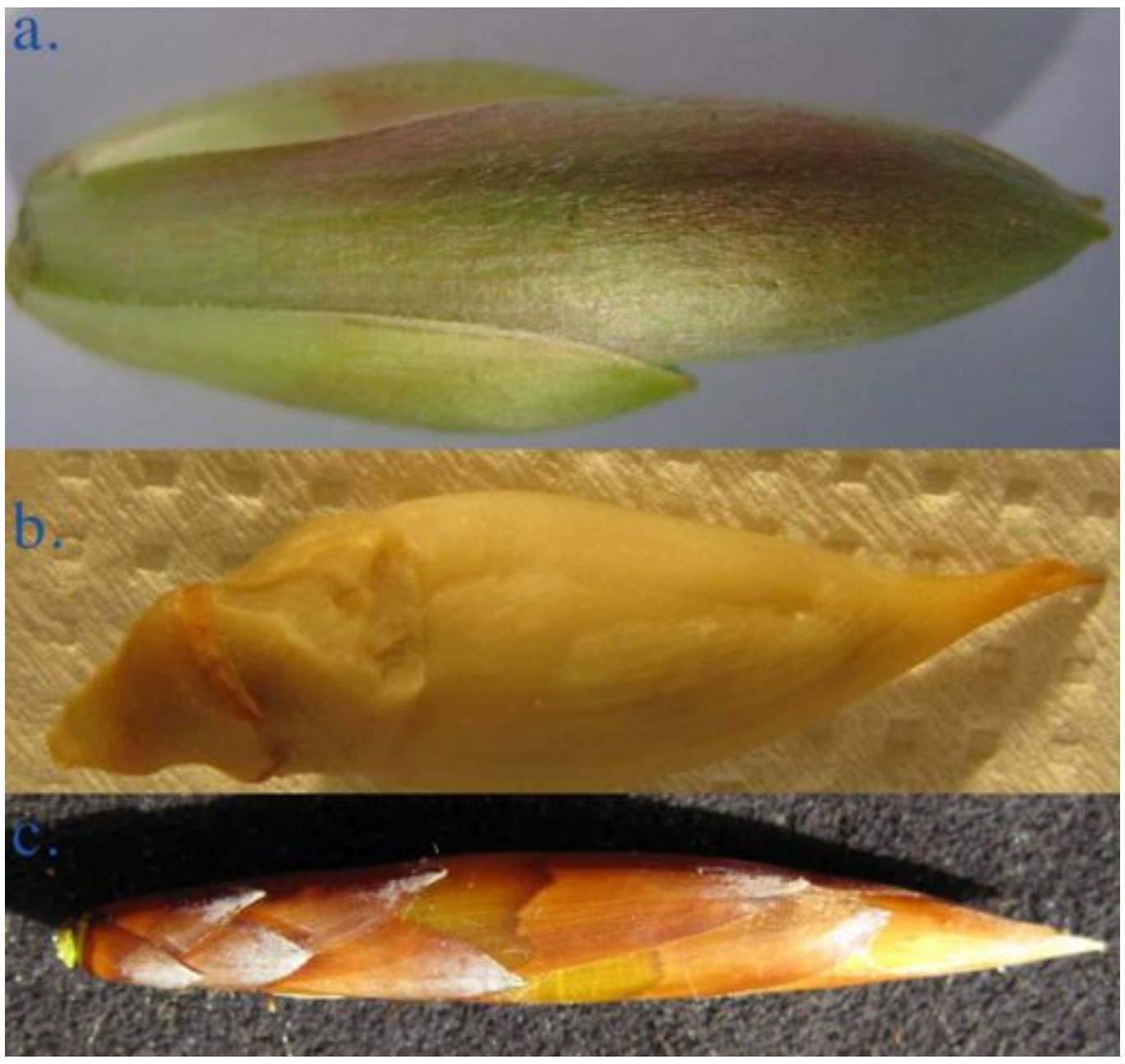

Figure 25. Different bud shape. a: *Acer pseudoplatanus*. b: *Ficus cariaca*. c: *Fagus sylvatica*. The bud b and c are moe acute than the first one.

## 4 - Conclusion

We have presented a panorama of the different kind of leaf packing and there consequences on leaf shape. We have shown how a mild modification of folding can have a drastic effect on leaf shape. The main conclusion of this work is that to guess the shape of a palmate leaf you need only to know the organisation of the folds and their orientation between them and toward the volume of the bud. By playing with these two parameters, you can create nearly all the leaves shape.

The fact that the bud is full (Filling law) needs that the lamina reaches the border of the constraining volume. Together with the folds, this induces a limitation of the lamina as if it has been cut to fit the volume (Kirigami Property). This in turn determines completely the unfolded shape. The fact that this law remains verified in such diverse geometries as presented here, shows that it is undoubtedly a key rule underlying leaf shape evolution. All the various leaves shape can be understood as different efficient way to fill the space at there disposal in the bud with different kind of folds

The folding stage and its packing regulation is necessary to bridge the gap between the primordia stages of development, which is being thoroughly studied, and the final shape of the leaves. The fact that the perimeter of these different folded lobes fall at the same border reveals a regulation process, aimed at filling perfectly the bud volume, and we propose it to be a mechanical contact regulation.

The regulation processes revealed by this leaf development study (the folding of the leaf around the veins, the mechanically sensitive leaf margin, and the overall flatness of the leaf), deserve to be more studied, and in particular their underlying molecular mechanisms. It would be also interesting to study in detail the evolutionary aspects to understand what are the evolutionary pressure which make the packing change.

## Annex – Material and numerical folding methods

### Material

*Acer pseudoplatanus* leaves were picked either in October either in June in different gardens and forest in and around Paris, France. *Tetrapanax papyriferum* leaves come from our lab specimens, Pheonix Botanical garden in Nice, and Val Rahmeh Botanical garden in Menton. Buds of *Fagus sylvatica*, *Fagus sylvatica Rohan obelix* and *Ficus cariaca* were collected in late spring in the Arboretum of Joinville-le-pont.

### Numerical folding method

Leaves are numerically folded back using these drawing of their veins (synclinal folds), contour and anti-veins (anticlinal folds, fig. 26 a). For instance we take the picture of mature *Acer pseudoplatanus* or *Tetrapanax papyriferum* leaves. Then the main veins and secondary veins are drawn (by hand), acquiring the numerical positions. The contour is also acquired numerically.

The first main veins and anti-veins all join in the same point, at the end of the petiole. For a first refolding of the leaf, we just measure the angles between the successive first segment of veins and anti-veins (fig.26 b), and redraw these segments by inverting the sign of one angle on two.

To fold the rest of the vein (*i.e* after the first segment) we used two different methods, one for the central vein, which is nearly a straigth line, and one for the lateral curved one. For the central vein we need to take into account the secondary folds (Figure 26 Bc), and for the lateral one the curved antivein (Figure 29 a a' b b', see below).

After having folded these veins and antiveins, we reconstruct the contour as above. Before folding, we decompose the initial contour on segments joining two consecutive folds extremities and the normal at this segment (orientation). After folding, we recompose the contour putting each segment on the new position of the end of veins and antiveins.

- *Case of secondary folds*

To fold secondary folds there can be a geometrical problem. It is not always possible to keep the actual angles as resulting the folded sheet may not lie flat. The method used to numerically fold the leaves onto a plane keeps fold lengths and aims to keep at best the angle values between the folds. Finding close values is a way to project the folding into the plane (or look at it from the side), with minimal distortions. Lets consider the branching detail on figure 26e, sketched in figure 26f. When unfolded, a secondary vein is branching at a primary vein with an angle $\mu$. The primary vein is making an angle $\nu$ at this branching point. Between these two anticlinal folds, *i.e.* the secondary and the primary veins, stands a synclinal fold that makes angles $\alpha$ and $\beta$ with them. To these angles $\alpha$, $\beta$, $\mu$ and $\nu$ for the unfolded leaf, correspond the angles $\alpha'$, $\beta'$, $\mu'$ and $\nu'$ when the leaf is folded (fig. 26g).

The sum of these angles for the unfolded leaf is of course:

$\alpha + \beta + \nu + \mu = 2\pi$

Considering the angle $\nu'$, the sketch of figure 26g is in a plane only if (Kobayashi Theorem) :

$\nu' = \mu' - \alpha' + \beta'$ (1)

If the sum of folded angles also follows (still on a plane unfolded)

$\alpha' + \beta' + \nu' + \mu' = 2\pi$,

equation (1) rewrites:

$\nu' = \pi - \alpha'$.

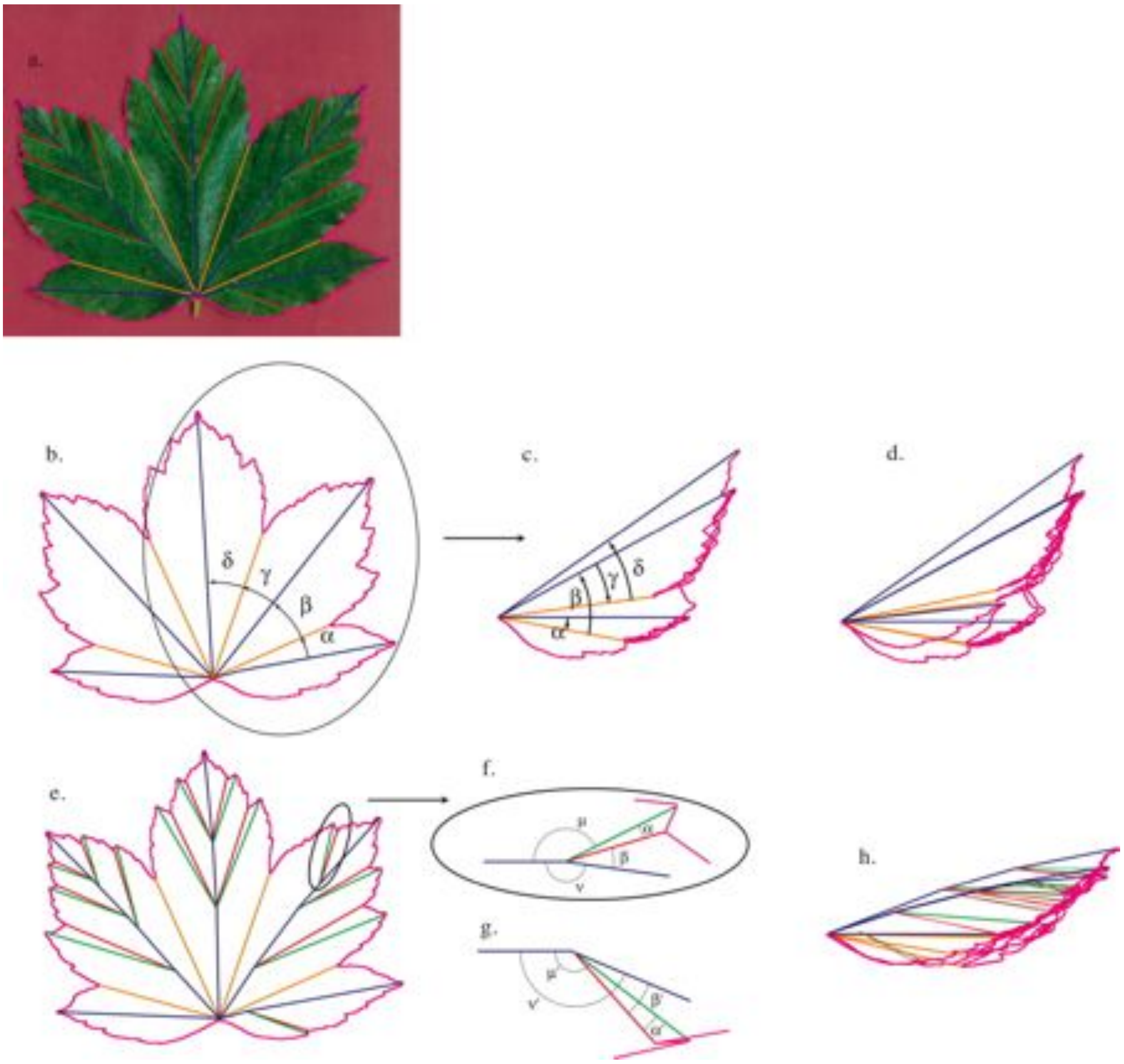

Figure 26 a. On a leaf, the first order vein (blue line) of a maple leaf and secondary vein (green line) are drawn. The first order anti-vein (orange line) which go from the intersection of two vein up to the sinus between these veins, and the second order anti-veins (red line) are also drawn. The contour of the leaf (mangenta line) is numerically detected. b. The result with the contour and only the main veins and anti-veins, and the first angles between them. c. Half part of the precedent sketch once refolded. $\alpha$ and $\gamma$ are reversed, and their respective contour drawn inversed. d. The refolded leaf using only its main folds. e. Sketch of the leaf with all its folds: secondary one and main one. f. Scheme of a secondary fold, unfolded. g. Folded in a plane. The news angles are obtained as described in the text. h. The whole set of veins and anti-veins is then drawn, with their respective contour, giving the completely refolded leaf.

Folding the branching with keeping to the best the angle $\nu$ is then minimizing the quantity:

$(\nu' - \nu)^2 + (\alpha' - \alpha)^2$

which rewrites

$(\nu' - \nu)^2 + (\pi - \nu' - \alpha)^2$.

One finds the best $\nu'$ value:

$\nu' = (\pi + \nu - \alpha) / 2$.

In the same way, one finds:

$\alpha' = (\pi + \alpha - \nu) / 2$,

$\beta' = (\pi + \beta - \mu) / 2$ and

$\mu' = (\pi + \mu - \beta) / 2$.

Once the angle corrected, the whole figure of folded veins and antiveins is drawn, and finally the contour is drawn for each vein-antivein segment, reverted if necessary, and with its angular position stretched or compressed if necessary, keeping the distance to the fold center. Usual corrections from the actual angles are around few degrees. For instance, in Figure 6 the corrections are around 1.5°.

- *Case of curved folds*

To refold the curved lateral lobes, we have taken their symmetric using the curved adjacent antifold as axis of symmetry (Figure 27a -b). For this the vein and antifold are cut in small elements, and each vein element is reflected around the closest antifold element. We then stretch if necessary the leaf perimeter to fit to the new length of the new vein. We made the assumption that the curvature due to the secondary folds is already taken in account in the curvature of the vein. We have located the secondary folds at their curviligne abscissa along the already refolded main vein (Figure 28).

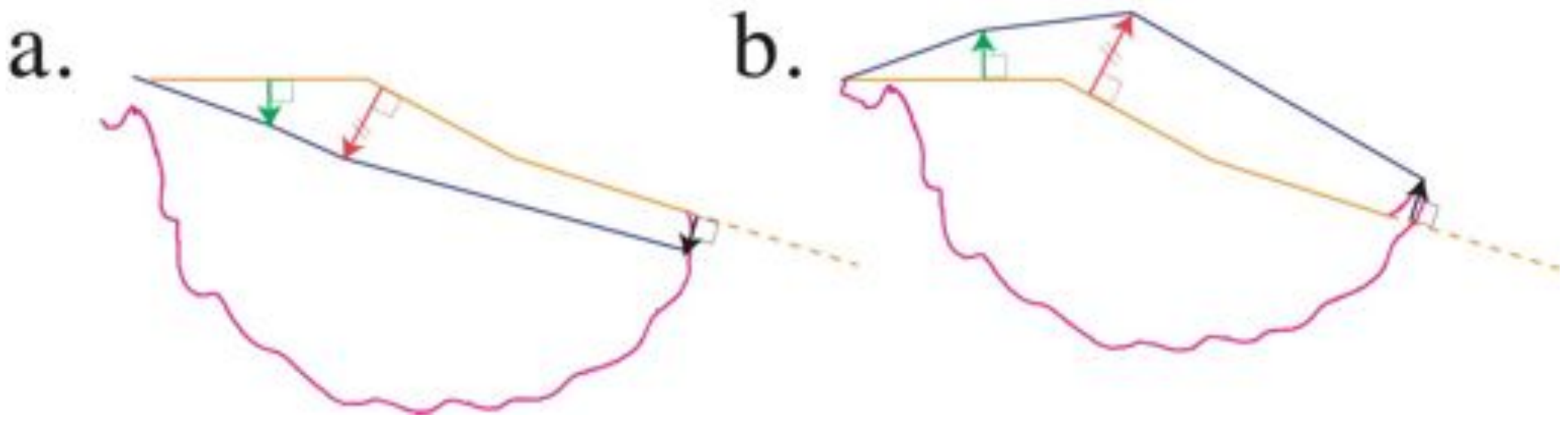

Figure 27: Symmetry around a curve fold. a. Antifold are drawn with an orange line, folds by a blue one, and the leaf margin by a magenta one. b. After the symmetry of the fold using the antifold as axis of symmetry.

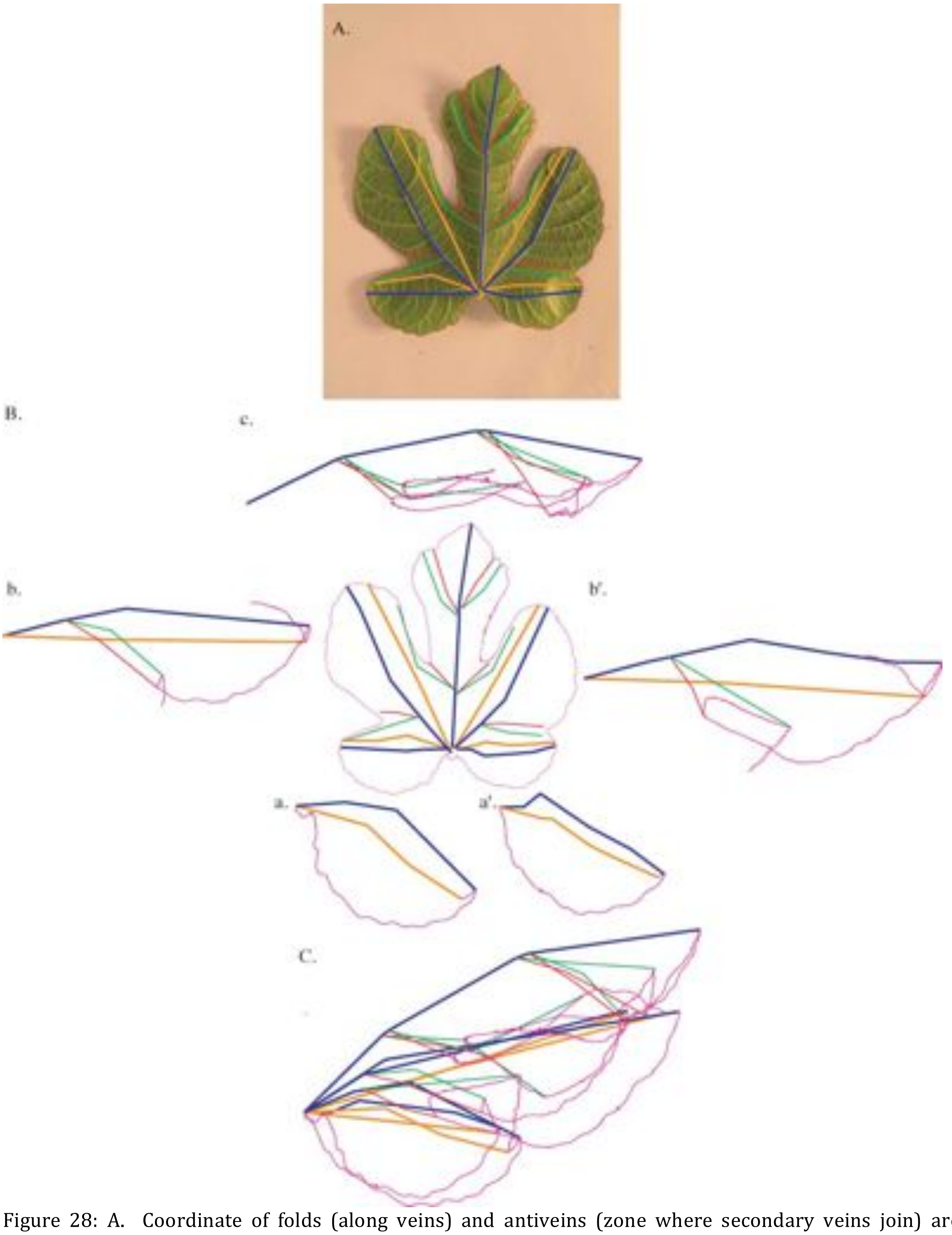

Figure 28: A. Coordinate of folds (along veins) and antiveins (zone where secondary veins join) are measured on a Ficus cariaca leaf image. B. Folding method. a. We orientate the left lateral lobe symetric (using the antifold as axias) in such way that the first segment of the fold is horizontal. b. We draw the following fold symetric turned with the angle between this fold and the precedent antifold. c. We represent the central lobe with its secondary folds. a'. b'. Idem of a. and b on the right side of the leaf. C. Final result.